\documentclass{iopart}

\usepackage{graphicx}
\usepackage{grffile}
\usepackage{subfigure}
\usepackage{color}
\usepackage{iopams}
\usepackage{setstack}
\usepackage{microtype}
\usepackage{multirow}

\begin{document}

\title[Numerical Continuation of  Bound \& Resonant States]{Numerical Continuation of Bound and Resonant States of the Two Channel Schr\"odinger Equation}
\author{P K\l{}osiewicz, W Vanroose and J Broeckhove}
\address{Department of Mathematics and Computer Science, University of Antwerp, Middelheimlaan 1, 2020 Antwerp, BE}
\eads{\mailto{przemyslaw.klosiewicz@ua.ac.be}, \mailto{wim.vanroose@ua.ac.be}, \mailto{jan.broeckhove@ua.ac.be}}

\begin{abstract}

Resonant solutions of the quantum Schr\"odinger equation occur at complex 
energies where the $S$-matrix becomes singular. Knowledge of such resonances 
is important in the 
study of the underlying physical system. Often the Schr\"odinger equation is
dependent on some parameter and one is interested in following the path of 
the resonances in the complex energy plane as the parameter changes. This is
particularly true in coupled channel systems where the resonant behavior is highly 
dependent on the strength of the channel coupling, the energy separation of the 
channels and other factors.

In previous work it was shown that numerical continuation, a 
technique familiar in the study of dynamical systems, can be brought to bear on the 
problem of following the resonance path in one dimensional problems \cite{Broeckhove2009} 
and multi-channel problems without energy separation between the 
channels \cite{klosiewicz2010}. A regularization can be defined that eliminates coalescing 
poles and zeros that appear in the $S$-matrix at the origin due to symmetries. Following the 
zeros of this regularized function then traces the resonance path.

In this work we show that this approach can be extended to channels with energy 
separation, albeit limited to two channels. The issue here is that the energy 
separation introduces branch cuts in the complex energy domain that need to be 
eliminated with a so-called uniformization. We demonstrate that the resulting 
approach is suitable for investigating resonances in two-channel systems and
provide an extensive example.

\end{abstract}

%
%
%
\submitto{\JPA}
\maketitle


\section{Introduction}
\label{sec:introduction}
The presence of resonances can drastically increase the yield of many
quantum mechanical reactions \cite{hotop2003}.  A resonance is an
intermediate state that is formed when reagents collide with an
appropriate energy and form an excited complex that decays into
reactants or products. In molecular reactions these intermediate
states are often electronic excited states. In the Born-Oppenheimer
picture the molecular dynamics temporarily follows the electronic
potential energy surface formed by the resonant state.

This picture forms the basis of state-of-the-art ab initio
calculations where, first, the position and lifetime of the resonance
is determined using electronic scattering calculations with the nuclei
fixed in space. This calculation needs to be repeated for every
possible configuration of the nuclear positions probed by the chemical
reactions and this results in the resonant potential energy surface.
In the second step, the nuclear dynamics is simulated on the resonant
potential energy surface that leads to the reactants. This approach
has been successfully used  to calculate the yields of processes such as
dissociative electron attachment to the water molecule
\cite{haxton2004,haxton2004a,haxton2007} and vibrational excitation of
carbon dioxide \cite{mccurdy2003,vanroose2004}, to name a few of the
processes that are mediated by a resonant state.

This work focuses on the first step outlined above, where the potential curves are
calculated as a function of the nuclear degrees of freedom.  We will
study a model consisting of a coupled Schr\"odinger equation and
construct, in an automatic way, the potential curve of a resonance
that becomes bound as the parameters in the equation are varied.  The
method tracks the resonances accurately, even in parameter
ranges where the resonance is too broad to be tracked with traditional
methods such as complex scaling \cite{moiseyev1998quantum}.

In this article, as is regularly done in the literature, a resonance
energy is defined as the complex energy at which the $S$-matrix has a
pole \cite{Newton1982,Taylor2006}. The real part of the pole position
defines the physical resonance energy and the imaginary part defines
the resonance decay width (or inverse lifetime).  The advantage of
this approach is that a bound state energy is also a pole of the
$S$-matrix, albeit with zero decay width i.e.\ infinite lifetime.  As
such, bound states and resonances are both treated in the same way
which makes it more convenient to trace resonant states as they become
bound, or vice-versa, when a problem parameter changes. The
  poles of the $S$-matrix will be found by a Newton iteration applied to a function that is proportional to the Jost function.

We aim to trace these states with numerical continuation. It is a
technique that has found widespread application especially in the
dynamical systems community. Assume one is given a solution $(u_0,\lambda_0)$ of
a set of $n$ nonlinear equations
$F(u,\lambda)=0$, where $F:\mathbb{R}^{n+1}\rightarrow\mathbb{R}^{n}$.
A second solution
$(u_1,\lambda_1)$ is then constructed numerically by applying a
predictor-corrector scheme. Repeated application constructs an
approximation of the implicitly defined solution set of $F$. One of the
well-known numerical continuation techniques is pseudo-arclength
continuation proposed by H.~B.~Keller~\cite{Keller1977}. It has been implemented
in computer programs e.g.\ AUTO~\cite{Doedel1981,AUTO2007},
LOCA~\cite{Salinger2002} and other numerical continuation
libraries. Because the corrector step in these methods is based on
Newton iterations, the derivatives of the function $F$ should be
Lipschitz continuous to guarantee fast convergence.

A traditional method to find the resonance position and width is \emph{complex scaling}~\cite{moiseyev1998quantum}. In this method, one applies a complex
  scaling transformation, $r \rightarrow r^{i \theta}$, to the
  reaction coordinate. This turns the resonance wave function into a
  square integrable function.  After the transformation the resonant
  state is then part of the discrete spectrum of the
  Hamiltonian.  The method, and its exterior variant (ECS) which has the advantage that it leaves the interactions in the inner region unchanged, have been
  successfully applied to find resonances in molecular systems such as
  HCO \cite{ryaboy1995three}, NeICL \cite{lipkin1993three} and in many
  other examples in atomic, molecular and nuclear physics.

Complex scaling, however, has its limitations. First, in
  numerical calculations the resonance position depends slightly on
  the choice of the rotation angle $\theta$. This is documented for
  example in \cite{moiseyev1998quantum}.  A second limitation is that
  only resonant states in a limited region in the
  complex energy plane can be found, in particular, a pie slice between the
  continuum spectrum, which is rotated $2\theta$ downwards from the real
  axis, and the real axis.  Virtual states for example, have a
  purely imaginary wave number and lie outside this region. As such they are hard to
  find with this method.

A resonance often transforms, very shortly, into a virtual state before it
  becomes a bound state as the parameters of the system change.
  To understand how resonances and bound states are connected through these states it is
  necessary to have a mathematical description that can handle these virtual states.

This paper investigates the application of numerical continuation
  to trace resonant states in a
  coupled channel Schr\"odinger equation as the parameters of the
  problem change. Unfortunately, the application of numerical
continuation to trace resonant states is not without challenges. For
Newton's method to work efficiently it requires a function whose
derivatives are Lipschitz continuous \cite{Kelley1995}. The $S$-matrix
does not satisfy these smoothness conditions.  In particular when a
resonance makes the transition to a bound state, a pole and a zero of
the $S$-matrix meet and straightforward application of numerical
continuation fails to trace the zero of $1/S$.

In \cite{Broeckhove2009} it was found, for one-dimensional quantum systems, 
that it is possible to apply the continuation to a regularized
function derived from the $S$-matrix because that function does satisfy the necessary smoothness conditions.
For several realistic potentials the resonances were tracked as parameters in the system were varied.  

In \cite{klosiewicz2010} the method has been extended to one-dimensional, many-channel 
problems where all channels have the same asymptotic energy threshold. Applications 
have demonstrated the viability of the approach in tracing the parameter dependence 
of the resonance in the system. However, as it stands, the method does not apply to 
systems where the channels have unequal energy thresholds. In this case,
branch cuts occur in the complex energy plane making the method invalid.

This is a significant restriction for certain application areas, e.g.\ most problems in molecular dynamics 
have such unequal energy thresholds. In this paper, we will investigate this difficulty and show 
that in the two-channel case it can be overcome by introducing an appropriate uniformization of the complex plane.
In addition we demonstrate the numerical practice of the method using an extensive example. 

We proceed as follows: In section~\ref{sec:cctise} we formulate the equations of interest in the context 
of quantum mechanical systems. Section~\ref{sec:smatrix} describes the so-called scattering matrix or 
$S$-matrix, its relation to resonances and bound states as well as the complex geometries that occur in 
the $S$-matrix of coupled channel systems. Section~\ref{sec:numcont} gives a brief overview of the 
background and applications of numerical continuation techniques. Section~\ref{sec:implementation} 
outlines the details of our implementation of the methods described in the preceding sections. 
In section~\ref{sec:results} we present an excerpt of the results we have obtained and compare with the ECS method and in
section~\ref{sec:conclusion} we give an outlook for possible applications and future studies.

\section{Coupled channel Schr\"odinger equation}
\label{sec:cctise}

The time independent Schr\"odinger equation for $N$ coupled channels with a spherically 
symmetric potential reads
\begin{equation}
\label{eq:154407092010}
	\left( -\frac{1}{2\mu}\frac{d^{2}}{dr^{2}}\mathbf{I} + \frac{\mathbf{L}(\mathbf{L}+\mathbf{I})}{2\mu r^{2}} + \mathbf{V}(r,\lambda) + \mathbf{\Xi} \right)\mathbf{\Psi}(r; E, \lambda) = E\mathbf{\Psi}(r; E,\lambda),
\end{equation}
where ($r\in\mathbb{R^{+}}$) is the radial coordinate, $\mathbf{I}$ is the $N \times N$ identity 
matrix, $\mathbf{\Xi}$ is the diagonal matrix of channel thresholds $\xi_{i}$, $\mathbf{L}$ 
is the diagonal matrix of channel angular momenta $l_{i}$, $\mathbf{V}(r,\lambda)$ is the matrix of 
channel and coupling potentials $V_{ij}(r,\lambda)$ which depend on a problem 
parameter $\lambda \in \mathbb{R}$ and, finally, $\mathbf{\Psi}(r)$ is the matrix of channel wave 
functions. Depending on the properties of the potential matrix $\mathbf{V}(r,\lambda)$ at infinity 
the behavior of the solutions $\mathbf{\Psi}(r)$ will differ significantly. We assume 
so called \emph{short range} interactions: $V_{ij}(r)$ vanishes faster than $r^{-3}$ as $r\to\infty$ and 
is less singular than $r^{-2}$ in the origin $r=0$, see also \cite{Taylor2006}.

Given homogeneous Dirichlet boundary conditions at $r=0$ and an incoming plane wave, the 
asymptotic ($r\to\infty$) solutions behave as:
\begin{equation}
\label{eq:114403112010}
	\mathbf{\Psi}(r; E,\lambda) = \frac{i}{2}\left[ \hat{\mathbf{h}}_{L}^{-}(\mathbf{K}r) - \hat{\mathbf{h}}_{L}^{+}(\mathbf{K}r) \mathbf{K}^{-\frac{1}{2}} \mathbf{S}(E,\lambda) \mathbf{K}^{\frac{1}{2}} \right],
\end{equation}
where $\mathbf{K} = \sqrt{2\mu(E\mathbf{I}-\mathbf{\Xi})}$ is the matrix of \emph{channel momenta} $k_{i} = \sqrt{2\mu(E-\xi_{i})}$ and
\begin{equation}
	\hat{\mathbf{h}}_{L}^{\pm}(\mathbf{K}r) =
	\left(\begin{array}{ccc}
		\hat{h}_{l_{1}}^{\pm}(k_{1}r) \\
		& \ddots \\
		&& \hat{h}_{l_{N}}^{\pm}(k_{N}r)
	\end{array}\right),
\end{equation}
is the matrix of spherical Riccati-Hankel functions associated with the various channels. The first 
term in (\ref{eq:114403112010}) is the partial wave expansion of the incoming plane wave, the second 
term represents, for each incoming partial wave, the outgoing wave in all the channels. The 
matrix $\mathbf{S}(E,\lambda)$ is the so-called scattering or $S$-matrix.

Our main object of interest is $\mathbf{S}(E,\lambda)$ because it contains all the information about the 
scattering process. It can be obtained from the coupled channel wave functions $\mathbf{\Psi}$ through the expression:
\begin{equation}
\label{eq:154607092010}
	\mathbf{S}(E,\lambda) = \mathbf{K}^{-\frac{1}{2}} \mathcal{W}\left[ \hat{\mathbf{h}}_{L}^{-}(\mathbf{K}r_{0}), \mathbf{\Psi}(r_{0}; E,\lambda) \right] {\mathcal{W}\left[ \hat{\mathbf{h}}_{L}^{+}(\mathbf{K}r_{0}), \mathbf{\Psi}(r_{0}; E,\lambda) \right]}^{-1} \mathbf{K}^{\frac{1}{2}},
\end{equation}
where the $\mathcal{W}$ stands for the Wronskian of two functions, whose usual definition 
\begin{equation}
	\mathcal{W}\left[f(x),g(x)\right] = f(x)\frac{dg(x)}{dx} - \frac{df(x)}{dx}g(x)
\end{equation}
is extended to matrices of functions as in \cite{Sitnikov2003}:
\begin{equation}
	\mathcal{W}\left[A(\mathbf{x}),B(\mathbf{x})\right] = A^{T}(\mathbf{x}) \frac{dB(\mathbf{x})}{d\mathbf{x}} - \frac{dA^{T}(\mathbf{x})}{d\mathbf{x}}B(\mathbf{x}),
\end{equation}
where $T$ is the transpose of a matrix, which vanishes in the specific case of expression~\eref{eq:154607092010}.

The Wronskians in equation~\eref{eq:154607092010} are 
evaluated at a point $r_{0}$ outside the range of 
the potentials, usually near the edge of the computational domain. Therefore, given a numerical method 
for solving the Schr\"odinger equation~\eref{eq:154407092010}, the $S$-matrix can be obtained numerically 
by evaluating matrix expression~\eref{eq:154607092010} which also involves the derivative of the wave 
functions. The details of our numerical implementation are 
given in section~\ref{sec:implementation}.


\section{Extracting the resonant and bound states from the $S$-matrix}
\label{sec:smatrix}

We are interested in solutions of equation~\eref{eq:154407092010} that correspond to the bound states and 
the resonances of the system. Many characterizations of these special solutions exist, yet the theoretically 
fundamental definition describes both bound and resonant states as an eigenstate of 
equation \eref{eq:154407092010} with \emph{purely outgoing} wave functions at infinity as the second boundary condition \cite{Newton1982,Taylor2006}.
 
A consistent alternative definition for multichannel systems interprets these states as having energies $E$ for 
which $\det(\mathbf{S}(E,\lambda))$ exhibits a pole.  
This is consistent with the first definition, as is easy to see when one
looks at the expression for the asymptotic wave
function \eref{eq:114403112010}. If $\det(\mathbf{S}(E,\lambda))$ has a pole, 
then in at least one eigenchannel (i.e.\ the channels defined by linear transformation
to diagonalize the $\mathbf{S}(E,\lambda)$ matrix) a resonant solution proportional 
to a purely outgoing wave occurs. A more formal approach relies on the introduction
of Jost functions, for which we refer to the literature \cite{Newton1982,Taylor2006}.

From the definition of the bound and the resonant states as a pole of
$\det(\mathbf{S}(E,\lambda))$ it is now possible to study the evolution of these
states in terms of a changing system parameter $\lambda\in\mathbb{R}$.
As $\lambda$ traverses the parameter space,
the bound and resonant states change position and lifetime. It is of
interest for many applications to know the explicit dependence on the
parameter of choice because it leads to the resonant and bound state potential
surface. 

It would be straightforward to define resonance trajectories
as curves in the complex $E$-plane parameterized by $\lambda$ i.e.\ $\left\{ (E,\lambda) \in\mathbb{C}\times\mathbb{R} \ \big| \ \det(\mathbf{S}(E,\lambda))^{-1}=0 \right\}$. Although this is theoretically
correct, it is not feasible numerically because of two reasons. First, in
the general case, the coupled channel $S$-matrix is a multi-valued
complex function of the energy and ``lives'' on a multi-sheeted
Riemann surface with branch cuts for every threshold value.  Second,
near threshold parameter values where bound and resonant states meet,
multiple poles and zeros of the $S$-matrix coalesce, thereby
destroying local smoothness properties. This is a consequence of well known 
symmetry properties of the $S$-matrix  \cite{Newton1982,Taylor2006}.

The presence of branch cuts and the coalescence of zeros and poles 
have a strong negative impact on the convergence of the Newton iteration 
used in solving
$\det(\mathbf{S}(E,\lambda))^{-1}=0$ when calculating the resonance trajectory.
A study of specific cases will give us insight in how these issues can be addressed.

\subsection{The case of $N$ channels, equal thresholds}
This case has been investigated in \cite{klosiewicz2010} and we briefly review the results.
In eq.~\eref{eq:154407092010} the matrix of channel thresholds $\mathbf{\Xi}$ is zero (i.e.\ we take the threshold energy to be zero in all channels). Consequently, the matrix of channel momenta $\mathbf{K}$ becomes
$k\mathbf{I}$ with $k=\sqrt{2\mu E}$. This simplifies expression \eref{eq:114403112010} to
\begin{equation}
\label{eq:153704112010}
	\mathbf{\Psi}(r;E,\lambda) = \frac{i}{2}\left[ \hat{\mathbf{h}}_{L}^{-}(kr) - \hat{\mathbf{h}}_{L}^{+}(kr) \mathbf{S}(E, \lambda) \right],
\end{equation}
such that \eref{eq:154607092010} becomes
\begin{equation}
\label{eq:154003102010}
	\mathbf{S}(E,\lambda) = \mathcal{W}\left[ \hat{\mathbf{h}}_{L}^{-}(kr_{0}), \mathbf{\Psi}(r_{0};E,\lambda) \right] {\mathcal{W}\left[ \hat{\mathbf{h}}_{L}^{+}(kr_{0}), \mathbf{\Psi}(r_{0};E,\lambda)\right]}^{-1}.
\end{equation}
Because $k=\sqrt{2\mu E}$, the function $\mathbf{S}(E, \lambda)$ is defined on a two-sheet Riemann 
surface with a branch cut on the negative real axis $E\in\mathbb{R}^{-}$.
This destroys continuity of the $S$-matrix near the branch cut and makes numerical 
continuation difficult. Fortunately, in this case, one can express the $S$-matrix in 
terms of $k$ by introducing  $\mathbf{S}(k,\lambda)$, a single-valued function of $k$. Afterwards, the 
results can be easily translated back to the complex $E$-plane with $E=\frac{k^{2}}{2\mu}$. In mathematical 
terms such a procedure of reparametrization of a multi-valued complex function to a single-valued 
function is referred to as a \emph{uniformization}.

Another issue that has to be dealt with is the coalescence of poles and zeros near the $k=0$ threshold 
value. This can be circumvented by effecting the numerical continuation on the function
\begin{eqnarray}
	F(k,\lambda) &= \det\left( \mathcal{K}\left( \mathbf{S}(k,\lambda) - \mathbf{I} \right)^{-1} \right), \qquad (\mathcal{K})_{ii} = k^{2l_{i}+1} \\
		&= \left(\prod k^{2l_{i}+1}\right) \Big/ \det\left( \mathbf{S}(k,\lambda) - \mathbf{I} \right), \label{eq:120822122010}
\end{eqnarray}
instead of directly on the $S$-matrix. In \cite{klosiewicz2010} it is shown that this procedure does not introduce 
false solutions, i.e.\ the zeros of $F$ are precisely the poles of $\mathbf{S}$, and that it eliminates any 
singularities at $k=0$. This is an extension of a similar result for the one-dimensional, single-channel 
systems case \cite{Vanroose2009,Broeckhove2009}.

\subsection{The case of two channels, unequal thresholds}
\begin{figure}
	\centering
	\subfigure[With unequal thresholds, branch cuts at $(-\infty,\xi_{i}{]}$.]{\includegraphics[width=4cm]{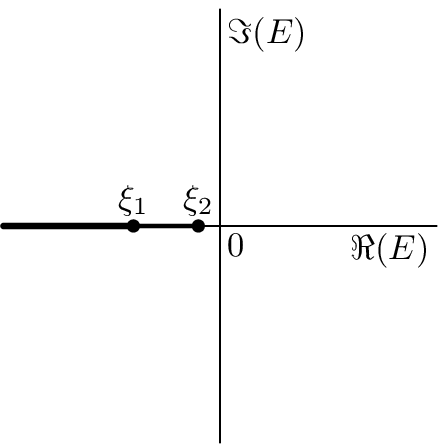}\label{fig:bc_two_channel_e}}
	\hspace{0.3cm}
	\subfigure[Even in the $k_{1}$ or $k_{2}$-plane there is a branch cut at {$[-b,b]$}]{\includegraphics[width=4cm]{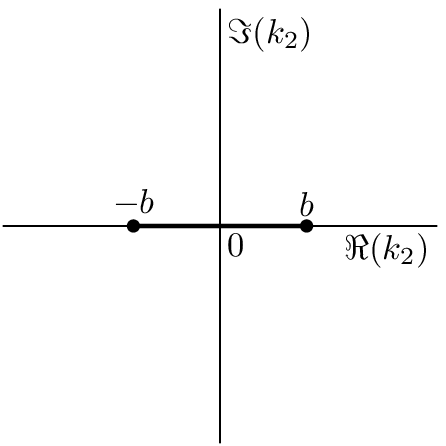}\label{fig:bc_two_channel_k2}}
	\caption{Branch cuts of $\mathbf{S}$ in $E$- and $k$-representation for a two channel system.}
	\label{fig:branchcuts}
\end{figure}

To highlight the difficulties that arise in the case of unequal thresholds we focus on a two channel model. Each threshold $\xi_{i}$ introduces a separate 
branch cut $(-\infty,\xi_{i}]$, see figure~\ref{fig:bc_two_channel_e}. Reparametrization 
of $\mathbf{S}$ in terms of either of the channel momenta $k_{i}$ does not provide a viable 
uniformization.  In each of the $k_{1}$ or $k_{2}$-planes
there is a branch cut at {$[-b,b]$} where $b=\sqrt{2\mu(\xi_{2}-\xi_{1})}$ (without loss 
of generality we take $\xi_{1} < \xi_{2}$), as illustrated in figure~\ref{fig:bc_two_channel_k2}. The 
cut disappears only when the thresholds coincide.  

For two channel systems a uniformization exists, given in~\cite{Newton1982} and modified 
slightly in~\cite{Sitnikov2003}. One introduces $u\in\mathbb{C}$ such that
\begin{equation}
	E(u) = \frac{\xi_{1} + \xi_{2}}{2} - \left|\frac{\xi_{2} - \xi_{1}}{2}\right| \frac{1+u^{4}}{2u^{2}}. \label{eq:142109122010}
\end{equation}
The corresponding expressions for channel momenta $k_{1}$ and $k_{2}$ are:
\begin{eqnarray}
	k_{1}(u) &= i\sqrt{\mu\frac{\xi_{2}-\xi_{1}}{2}}\frac{u^{2}-1}{u}, \\
	k_{2}(u) &= i\sqrt{\mu\frac{\xi_{2}-\xi_{1}}{2}}\frac{u^{2}+1}{u},
\end{eqnarray}
and the $2\times2$ $S$-matrix is now explicitly written as
\begin{equation}
	\mathbf{S}(u,\lambda) = \left(\begin{array}{cc}k_{1}(u) & 0 \\ 0 & k_{2}(u) \end{array}\right)^{-\frac{1}{2}}
			\mathbf{W}_{-}(u,\lambda)\left[\mathbf{W}_{+}(u,\lambda)\right]^{-1} \left(\begin{array}{cc}k_{1}(u) & 0 \\ 0 & k_{2}(u) \end{array}\right)^{\frac{1}{2}}, \label{eq:113419112010}
\end{equation}
where
\begin{eqnarray}
	\mathbf{W}_{\pm}(u,\lambda) &=  \left(\begin{array}{cc} \hat{h}_{l_{1}}^{\pm}(k_{1}(u)r_{0}) & 0 \\ 0 & \hat{h}_{l_{2}}^{\pm}(k_{2}(u)r_{0}) \end{array}\right) \left(\begin{array}{cc} \frac{d}{dr}\psi_{11}(r_{0};E(u),\lambda) & \frac{d}{dr}\psi_{12}(r_{0};E(u),\lambda) \\ \frac{d}{dr}\psi_{21}(r_{0};E(u),\lambda) & \frac{d}{dr}\psi_{22}(r_{0};E(u),\lambda) \end{array}\right) \\ &\quad- \left(\begin{array}{cc} \frac{d}{dr}\hat{h}_{l_{1}}^{\pm}(k_{1}(u)r_{0}) & 0 \\ 0 & \frac{d}{dr}\hat{h}_{l_{2}}^{\pm}(k_{2}(u)r_{0}) \end{array}\right) \left(\begin{array}{cc} \psi_{11}(r_{0};E(u),\lambda) & \psi_{12}(r_{0};E(u),\lambda) \\ \psi_{21}(r_{0};E(u),\lambda) & \psi_{22}(r_{0};E(u),\lambda) \end{array}\right),
\end{eqnarray}
where $\psi_{ij}(r_{0};E(u),\lambda)$ stands for the $(i,j)$-th element of the matrix 
wave function $\mathbf{\Psi}$ calculated for a specific parameter $\lambda$ and evaluated at a point $r_{0}$.

\begin{figure}
	\centering
	\includegraphics[width=6cm]{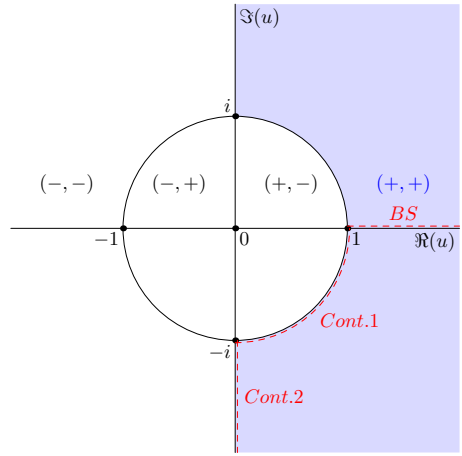}
	\caption{Uniformization of the four-sheeted Riemann surface of $\mathbf{S}(E)$ using the 
		transformation \eref{eq:142109122010} unravels the sheets onto four different 
		regions of the complex $u$-plane. The blue region $(+,+)$ corresponds to the 
		physical sheet $|u|>1, \Re(u)>0$. Bound states ({\color{red}$BS$}) $E\in(-\infty,\xi_{1}]$ lie on $u\in[1,+\infty)$. Scattering states ({\color{red}$Cont.1$}) $E\in[\xi_{1},\xi_{2}]$ map to $u=e^{i\theta}, \theta\in[-\frac{\pi}{2},0]$. Scattering states ({\color{red}$Cont.2$}) $E\in[\xi_{2},+\infty)$ map to $u\in[-i,-i\infty)$.}
	\label{fig:u_plane}
\end{figure}

The effect of this uniformization can be visualized by looking at the different parts of the $u$-plane and how they map on the $E$-plane as illustrated in \fref{fig:u_plane}. Four regions $(\pm,\pm)$ in the $u$-plane can be distinguished. They are separated by the imaginary axis and the unit circle and correspond to the four different sheets of $\mathbf{S}(E,\lambda)$ in the $E$-plane. The region labeled as $(+,+)$ bounded by $|u|>1$ and $\Re(u)>0$ is mapped to the physical sheet. Physical bound states $E\in(-\infty,\xi_{1}]$ are located on $u\in[1,+\infty)$. The physical continuum $E\in[\xi_{1},\xi_{2}]$ maps to the quarter unit circle in the fourth quadrant $u=e^{i\theta}, \theta\in[-\frac{\pi}{2},0]$; whereas the physical continuum $E\in[\xi_{2},+\infty)$ is mapped to $u\in[-i,-i\infty)$. A more detailed description of the different regions in the $u$-plane can be found in~\cite{Sitnikov2003,Newton1982}.

Continuation paths can traverse different regions in the $u$-plane, and as such, different sheets in the complex energy plane. The focus of this work is concentrated on finding these trajectories, independently of their precise physical meaning. The results we obtain are presented in the $u$-plane. However, they can be translated to the complex energy plane if one is interested in physically observable quantities.

Note that this uniformization is strictly limited to a two channel
case. A similar procedure for three channel systems is much more
involved, as indicated in \cite{Newton1982}. To the knowledge of the
authors there is no generalization for $N$ channels. For a thorough discussion of the analytical properties of the $S$-matrix and related functions in many-channel problems we refer to~\cite{Newton1982,Taylor2006}.

Having established a feasible way of extracting the $S$-matrix for a 
specific $u\in\mathbb{C}$, we are still faced with the problem of coalescing 
poles and zeros as discussed in the previous case of equal thresholds. Fortunately 
a straightforward extension of the regularization procedure~\eref{eq:120822122010} 
can be applied. Similarly, we create the function
\begin{eqnarray}
	F(u,\lambda) &= \det\left( \mathcal{K}\left( \mathbf{S}(u,\lambda) - \mathbf{I} \right)^{-1} \right), \qquad (\mathcal{K})_{ii} = k_{i}^{2l_{i}+1}(u) \\
		&= \left(\prod_{i=1}^{N} k_{i}^{2l_{i}+1}(u)\right) \Big/ \det\left( \mathbf{S}(u,\lambda) - \mathbf{I} \right),
\end{eqnarray}
which in our case reduces to
\begin{equation}
	F(u,\lambda) = \frac{k_{1}^{2l_{1}+1}(u) \ k_{2}^{2l_{2}+1}(u) }{ \det\left( \mathbf{S}(u,\lambda) - \mathbf{I}_{2} \right)}. \label{eq:142022122010}
\end{equation}
This is the function whose solution set will be approximated through numerical continuation to obtain the trajectories of the resonant $S$-matrix poles.

\section{Numerical continuation of resonances}
\label{sec:numcont}
In the previous section we have identified a resonance trajectory starting at $(E_{0}, \lambda_{0})$ as the implicitly defined curve $E=E(\lambda)$ with
\begin{equation}
F(E(\lambda),\lambda) = 0, \quad E(\lambda_{0}) = E_{0},
\end{equation}
where the function $F$ is defined in equation \eref{eq:142022122010}.
We are now interested in constructing such trajectories automatically and robustly. 

In dynamical systems similar equations arise in the study of steady states 
of parameterized ODEs and efficient methods have 
been developed to find the solution curves in terms of varying parameter values.
In this context, one is interested in finding the solution of an under determined system 
of nonlinear equations
\begin{equation}
	\label{eq:F}
	F:\mathbb{R}^{n+1} \longrightarrow \mathbb{R}^{n} \ : \ \mathbf{x}=(\mathbf{u},\lambda) \longmapsto F(\mathbf{x}),
\end{equation}
connected to an initial point  $\mathbf{x}_{0} = (\mathbf{u}_{0},\lambda_{0})$.

Many of these problems are computationally intensive and efficiency is a key concern in the numerical studies.
In particular, the number of evaluations of the function $F$ should be kept to a minimum. In addition,
the solution components often have complex geometries with intersections and 
bifurcations. The study of bifurcations generally involves rigorous stability analysis of the underlying solutions 
and is a complicated subject on its own. In this work only the so called ``simple'' bifurcation points can occur. 
These manifest themselves as two intersecting solution 
branches and are characterized by the dimension of the null-space of 
the Jacobian $F_{\mathbf{x}}(\mathbf{x}_{t})$ being 2 in 
a point $\mathbf{x}_t$, which is called a ``branch point''. In many problems, however, 
the bifurcations are much more involved and 
a thorough treatise on bifurcations and stability of solutions 
can be found in~\cite{Keller1977,Seydel1994,Doedel2007,Allgower1990}.

Numerical continuation is the process of solving the equation \eref{eq:F} by constructing 
successive approximate solutions on the path starting at the known solution $\mathbf{x}_{0}$. A good 
overview of these techniques can be found in \cite{Allgower1990}. There are essentially two different approaches to
tracing such paths viz.\ piece-wise linear methods and predictor-corrector methods. We will use one 
of the latter methods. Typically, one first makes a predictor step (Euler prediction is commonly used) 
that estimates the next point by following 
the tangent $\mathbf{t}_{i}$ to the curve at the current point $\mathbf{x}_{i}$ for a certain small 
distance $\Delta s$ as in $\mathbf{x}_{i+1}^{p} = \mathbf{x}_{i} + \Delta s\,\mathbf{t}_{i}$.
Next, a corrector step is applied to converge to a solution $\mathbf{x}_{i+1}^{p} \longrightarrow \mathbf{x}_{i+1}$. 
Quite often Newton iterations are used as a corrector.

\begin{figure}
	\centering
	\includegraphics[width=4cm]{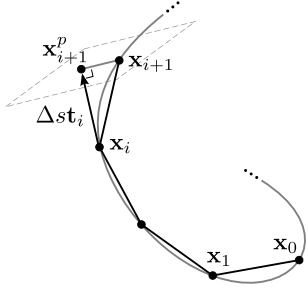}
	\caption{Schematic representation of pseudo-arclength continuation.}
	\label{fig:pac}
\end{figure}

One such predictor-corrector method, the one we will use in this paper, is the pseudo-arclength continuation.
Its corrector step consists of Newton iterations on the 
system $F(\mathbf{x})=\mathbf{0}$ augmented with an additional equation that constrains the 
iterations to a hyper plane through $\mathbf{x}_{i+1}^{p}$ and perpendicular to the 
tangent $\mathbf{t}_{i}$ thereby giving the next point $\mathbf{x}_{i+1}$ on the solution curve (see \fref{fig:pac}).

A robust implementation of numerical continuation that can detect branch points and continue 
branching solution curves is provided by AUTO~\cite{Doedel1981,AUTO2007}, which we have used in this work. 
Other well-known implementations of numerical continuation algorithms include LOCA~\cite{Salinger2002} which is a part of the Trilinos framework~\cite{Heroux2005}, \textsc{Matcont}~\cite{Dhooge2006} (a Matlab implementation) and Multifario~\cite{Henderson2002} which allows multi-pa\-ra\-me\-ter continuation.

\section{Numerical implementation}
\label{sec:implementation}
Due to its ability to automatically find and follow bifurcating branches, we have opted to use AUTO 
for numerical continuation. Therefore the main component that has to be provided to implement 
methods described in previous sections is the actual code that calculates the numerical values 
of the function $F(u,\lambda)$, see eq.~\eref{eq:142022122010}, for a given $u\in\mathbb{C}\cong\mathbb{R}^{2}$ and $\lambda\in\mathbb{R}$.

We have solved equation \eref{eq:154407092010} both with a $\mathcal{O}(h^{5})$ renormalized 
Numerov method detailed in \cite{Johnson1978,Johnson1977}. 
We also need the derivative of the numerical wave function at the end of the computational 
domain for a specified energy $E(u)$ in order to calculate the Wronskians in \eref{eq:113419112010}. 
We compute this derivative using a technique also described in \cite{Johnson1977} and which is of order $\mathcal{O}(h^{4})$. Finally, we need to calculate the function $F$ given by expression~\eref{eq:142022122010}.

We emphasize that the numerical continuation results are independent of the underlying numerical 
solver of the Schr\"odinger equation and that both $\mathcal{O}(h^5)$ and the $\mathcal{O}(h^2)$ methods 
give the same qualitative results. However, we have found that higher order methods for both the wave function 
and its derivative lead to significantly more robust continuation curves and allow for wider energy 
ranges. For this reason we selected the higher order renormalized Numerov method to generate 
results of section~\ref{sec:results}.

This approach has been implemented in C++ and is used as a driver routine by the AUTO program.

\section{Examples and results}
\label{sec:results}
As an example of the complicated geometries this method is able to cope with, we consider a coupled channel $s$-wave ($l_{1}=l_{2}=0$) system with Gaussian potential wells both as the channel potentials and the coupling. The $2\times2$ potential matrix has the elements:
\begin{equation}
\eqalign{
	V_{ii}(r,\lambda_{i}) &= -\lambda_{i} e^{-\frac{r^{2}}{4}} \\
	V_{i\neq j}(r,\lambda_{c}) &= \lambda_{c} e^{-r^{2}},
}
\end{equation}
where $\lambda_{1}$, $\lambda_{2}$ and $\lambda_{c}$ denote the potential strength of the first, second and coupling channels respectively. The channel thresholds are chosen $\xi_{1}=0$ and $\xi_{2}=\frac{1}{2}$ and the mass is $\mu=1$. Equation~\eref{eq:154407092010} was solved using the previously mentioned renormalized Numerov method on the domain $r\in[0,4.8]$ with 4096 grid points. All of $\lambda_{1}$, $\lambda_{2}$ and $\lambda_{c}$ will be used as variable system parameters. We will indicate clearly which of those are fixed and which are used as continuation parameters. 

\begin{table}
	\footnotesize
	\centering
	\begin{tabular}{|c|r|c|rr|rr|}
		\hline
		$\lambda_{c}$ & State & label & $\Re(u)$ & $\Im(u)$ & $\Re(E)$ & $\Im(E)$ \\
		\hline
		\multirow{8}{*}{\texttt{0}}
		&ch 1, $n_{0}$ & {\color{blue}$c_{1}\tilde{n}_{0}$} & \texttt{-2.2983975e-01} & \texttt{0} & \texttt{-2.1228484e+00} & \texttt{0} \\
		&& {\color{blue}$c_{1}n_{0}$} & \texttt{4.3508575e+00} & \texttt{0} & \texttt{-2.1228484e+00} & \texttt{0} \\
		&$n_{1}$ & {\color{red}$c_{1}\tilde{n}_{1}$} & \texttt{-4.5199837e-01} & \texttt{0} & \texttt{-3.8737558e-01} & \texttt{0} \\
		&& {\color{red}$c_{1}n_{1}$} & \texttt{2.2123974e+00} & \texttt{0} & \texttt{-3.8737558e-01} & \texttt{0} \\
		\cline{2-7}
		&ch 2, $n_{0}$ & {\color{green}$c_{2}\tilde{n}_{0}$} & \texttt{2.5892712e-01} & \texttt{0} & \texttt{-1.6228484e+00} & \texttt{0} \\
		&& {\color{green}$c_{2}n_{0}$} & \texttt{3.8620906e+00} & \texttt{0} & \texttt{-1.6228484e+00} & \texttt{0} \\
		&$n_{1}$ & {\color{magenta}$c_{2}\tilde{n}_{1}$} & \texttt{8.8019950e-01} & \texttt{4.7460388e-01} & \texttt{1.1262442e-01} & \texttt{0} \\
		&& {\color{magenta}$c_{2}n_{1}$} & \texttt{8.8019950e-01} & \texttt{-4.7460388e-01} & \texttt{1.1262442e-01} & \texttt{0} \\
		\hline
		$\lambda_{c}$ & State & label & $\Re(u)$ & $\Im(u)$ & $\Re(E)$ & $\Im(E)$ \\
		\hline
		\multirow{8}{*}{\texttt{0.2}}
		&ch 1, $n_{0}$ & {\color{blue}$c_{1}\tilde{n}_{0}$} & \texttt{-2.2923691e-01} & \texttt{0} & \texttt{-2.1352756e+00} & \texttt{0} \\
		&& {\color{blue}$c_{1}n_{0}$} & \texttt{4.3623083e+00} & \texttt{0} & \texttt{-2.1352854e+00} & \texttt{0} \\
		&$n_{1}$ & {\color{red}$c_{1}\tilde{n}_{1}$} & \texttt{-4.5179967e-01} & \texttt{0} & \texttt{-3.8789140e-01} & \texttt{0} \\
		&& {\color{red}$c_{1}n_{1}$} & \texttt{2.2141945e+00} & \texttt{0} & \texttt{-3.8832855e-01} & \texttt{0} \\
		\cline{2-7}
		&ch 2, $n_{0}$ & {\color{green}$c_{2}\tilde{n}_{0}$} & \texttt{2.5963744e-01} & \texttt{0} & \texttt{-1.6127068e+00} & \texttt{0} \\
		&& {\color{green}$c_{2}n_{0}$} & \texttt{3.8517883e+00} & \texttt{0} & \texttt{-1.6129594e+00} & \texttt{0} \\
		&$n_{1}$ & {\color{magenta}$c_{2}\tilde{n}_{1}$} & \texttt{8.7757633e-01} & \texttt{4.7363933e-01} & \texttt{1.1278823e-01} & \texttt{1.1579542e-03} \\
		&& {\color{magenta}$c_{2}n_{1}$} & \texttt{8.7757633e-01} & \texttt{-4.7363933e-01} & \texttt{1.1278823e-01} & \texttt{-1.1579542e-03} \\
		\hline
		$\lambda_{c}$ & State & label & $\Re(u)$ & $\Im(u)$ & $\Re(E)$ & $\Im(E)$ \\
		\hline
		\multirow{8}{*}{\texttt{0.3}}
		&ch 1, $n_{0}$ & {\color{blue}$c_{1}\tilde{n}_{0}$} & \texttt{-2.2852083e-01} & \texttt{0} & \texttt{-2.1501654e+00} & \texttt{0} \\
		&& {\color{blue}$c_{1}n_{0}$} & \texttt{4.3759865e+00} & \texttt{0} & \texttt{-2.1501849e+00} & \texttt{0} \\
		&$n_{1}$ & {\color{red}$c_{1}\tilde{n}_{1}$} & \texttt{-4.5155161e-01} & \texttt{0} & \texttt{-3.8853641e-01} & \texttt{0} \\
		&& {\color{red}$c_{1}n_{1}$} & \texttt{2.2164315e+00} & \texttt{0} & \texttt{-3.8951600e-01} & \texttt{0} \\
		\cline{2-7}
		&ch 2, $n_{0}$ & {\color{green}$c_{2}\tilde{n}_{0}$} & \texttt{2.6048642e-01} & \texttt{0} & \texttt{-1.6006947e+00} & \texttt{0} \\
		&& {\color{green}$c_{2}n_{0}$} & \texttt{3.8395472e+00} & \texttt{0} & \texttt{-1.6012444e+00} & \texttt{0} \\
		&$n_{1}$ & {\color{magenta}$c_{2}\tilde{n}_{1}$} & \texttt{8.7428785e-01} & \texttt{4.7241720e-01} & \texttt{1.1298423e-01} & \texttt{2.6183712e-03} \\
		&& {\color{magenta}$c_{2}n_{1}$} & \texttt{8.7428785e-01} & \texttt{-4.7241720e-01} & \texttt{1.1298423e-01} & \texttt{-2.6183712e-03} \\
		\hline
		$\lambda_{c}$ & State & label & $\Re(u)$ & $\Im(u)$ & $\Re(E)$ & $\Im(E)$ \\
		\hline
		\multirow{8}{*}{\texttt{0.5}}
		&ch 1, $n_{0}$ & {\color{blue}$c_{1}\tilde{n}_{0}$} & \texttt{-2.2645171e-01} & \texttt{0} & \texttt{-2.1939897e+00} & \texttt{0} \\
		&& {\color{blue}$c_{1}n_{0}$} & \texttt{4.4159879e+00} & \texttt{0} & \texttt{-2.1940286e+00} & \texttt{0} \\
		&$n_{1}$ & {\color{red}$c_{1}\tilde{n}_{1}$} & \texttt{-4.5076010e-01} & \texttt{0} & \texttt{-3.9060199e-01} & \texttt{0} \\
		&& {\color{red}$c_{1}n_{1}$} & \texttt{2.2235200e+00} & \texttt{0} & \texttt{-3.9328811e-01} & \texttt{0} \\
		\cline{2-7}
		&ch 2, $n_{0}$ & {\color{green}$c_{2}\tilde{n}_{0}$} & \texttt{2.6297322e-01} & \texttt{0} & \texttt{-1.5661803e+00} & \texttt{0} \\
		&& {\color{green}$c_{2}n_{0}$} & \texttt{3.8041557e+00} & \texttt{0} & \texttt{-1.5675876e+00} & \texttt{0} \\
		&$n_{1}$ & {\color{magenta}$c_{2}\tilde{n}_{1}$} & \texttt{8.6368879e-01} & \texttt{4.6837873e-01} & \texttt{1.1354314e-01} & \texttt{7.3933316e-03} \\
		&& {\color{magenta}$c_{2}n_{1}$} & \texttt{8.6368879e-01} & \texttt{-4.6837873e-01} & \texttt{1.1354314e-01} & \texttt{-7.3933316e-03} \\
		\hline
	\end{tabular}
	\caption{Numerical values of the states in the $u$- and $E$-planes. Different states with coupling values (top to bottom) $\lambda_{c}=0$, $\lambda_{c}=0.2$, $\lambda_{c}=0.3$ and $\lambda_{c}=0.5$ are shown. The colored labels are used in figures to make a clear distinction between states.}
	\label{tab:095115122010}
\end{table}

\begin{figure}[h!]
	\centering
	\subfigure[$\Re(u)\times\lambda_{c}$ projection]{%
		\includegraphics[width=9cm]{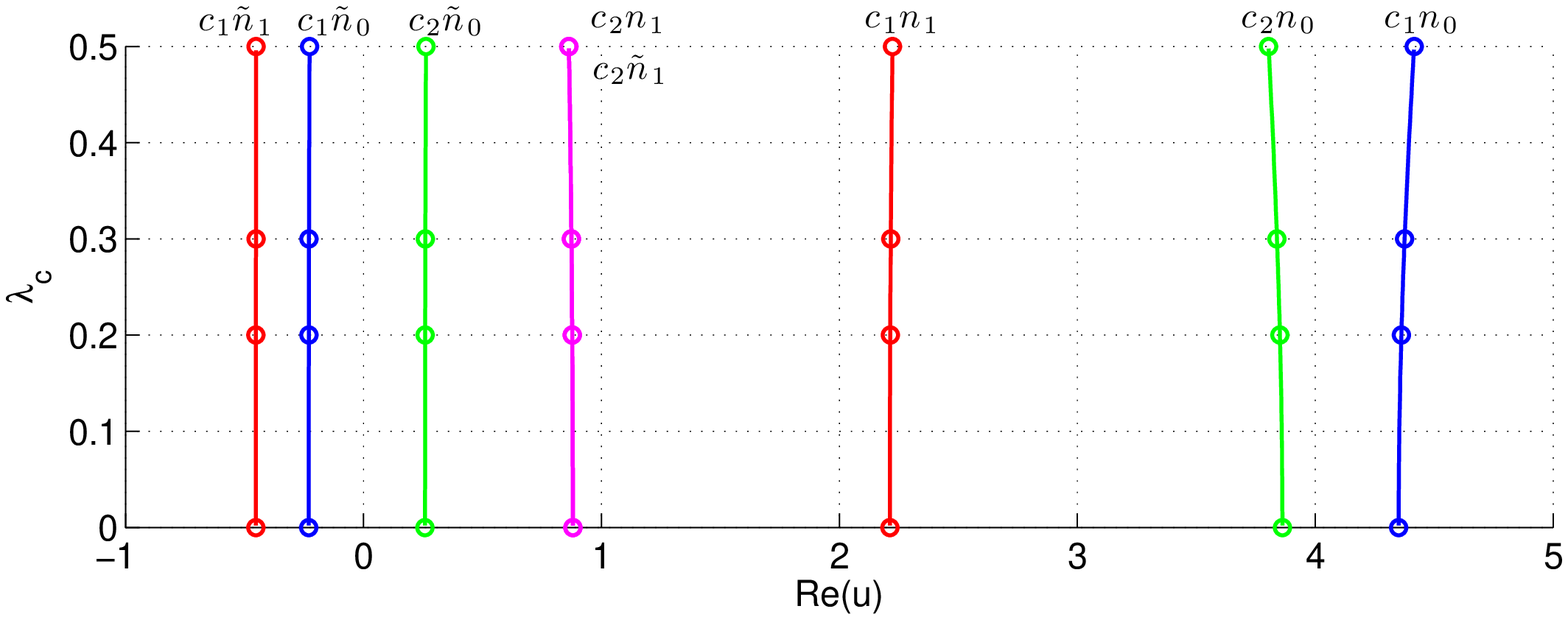}%
		\label{fig:in_coupling_re_u}%
	}
	\subfigure[$\Im(u)\times\lambda_{c}$ projection]{%
		\includegraphics[width=3cm]{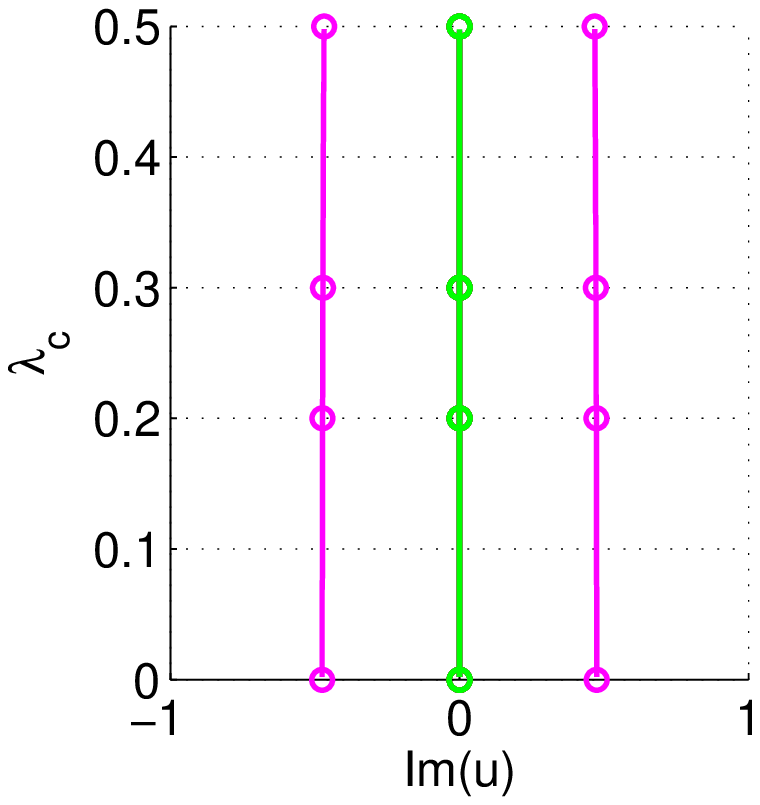}%
		\label{fig:in_coupling_im_u}%
	}
	\subfigure[$\Re(u)\times\Im(u)$ projection]{%
		\includegraphics[width=9cm]{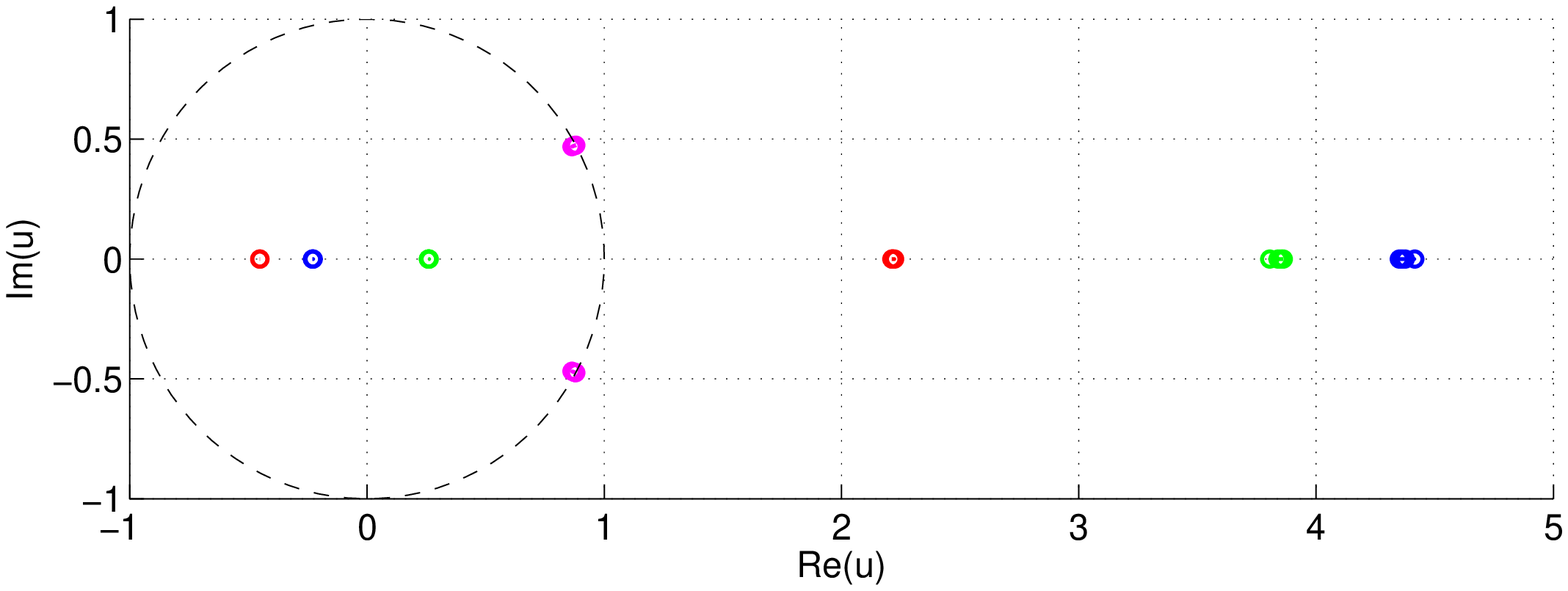}%
		\label{fig:in_coupling_reim_u}%
	}
	\hspace{3cm}
	\caption{(color online) Continuation paths of bound states of the example system in terms of increasing channel coupling strength $\lambda_{c}$ projected on three different planes. As the coupling strength increases, slight repulsion of the states in both channels can be observed. Circled values indicate starting points used for continuation in channel strengths $\lambda_{i}$. The values of these points are given in \tref{tab:095115122010}. In \fref{fig:in_coupling_reim_u} the dashed line represents the unit circle.}
	\label{fig:in_coupling}
\end{figure}

In the uncoupled case ($\lambda_{c}=0$), setting $\lambda_{i}=4$ gives a system with two bound states in each channel whose values can be found in the upper part of \tref{tab:095115122010}. Using these values as starting points we carry out the continuation in terms of increasing channel coupling $\lambda_{c}$ while keeping $\lambda_{i}=4$ fixed. The results in the $u$-plane are shown in \fref{fig:in_coupling}. Points corresponding to coupling values 0, 0.2, 0.3 and 0.5 are highlighted in the figure. Their corresponding numerical values in the $u$- and $E$-planes are summarized in \tref{tab:095115122010}. Notice that for $\lambda_{c}=0$, every value of $E$ is associated with two different points in the $u$-plane which are located on different sheets. Following the paths of such two values eventually gives rise to different values in the $E$-plane. Therefore, one needs to keep track of all of them. We give appropriate labels, see \tref{tab:095115122010}, to distinguish different points in the $u$-plane.

While continuing in $\lambda_{c}$, only slight variations in state's energy occur, yet the seemingly minor coupling has profound effects on the evolution of these states in terms of potential strengths $\lambda_{1}$ and $\lambda_{2}$ while keeping the coupling strength constant.

To illustrate this behavior we fix the coupling strength and use the corresponding $u$-values of the states as starting points for a continuation in terms of variations of both potential strengths $\lambda_{1}$ and $\lambda_{2}$ simultaneously. As $\lambda_{i}$ decreases, we expect these bound states to move into the resonant regime and influence each other due to the coupling.

\begin{figure}[h!]
	\centering
	\subfigure[$\Re(u)\times\lambda_{i}$]{%
		\includegraphics[width=8cm]{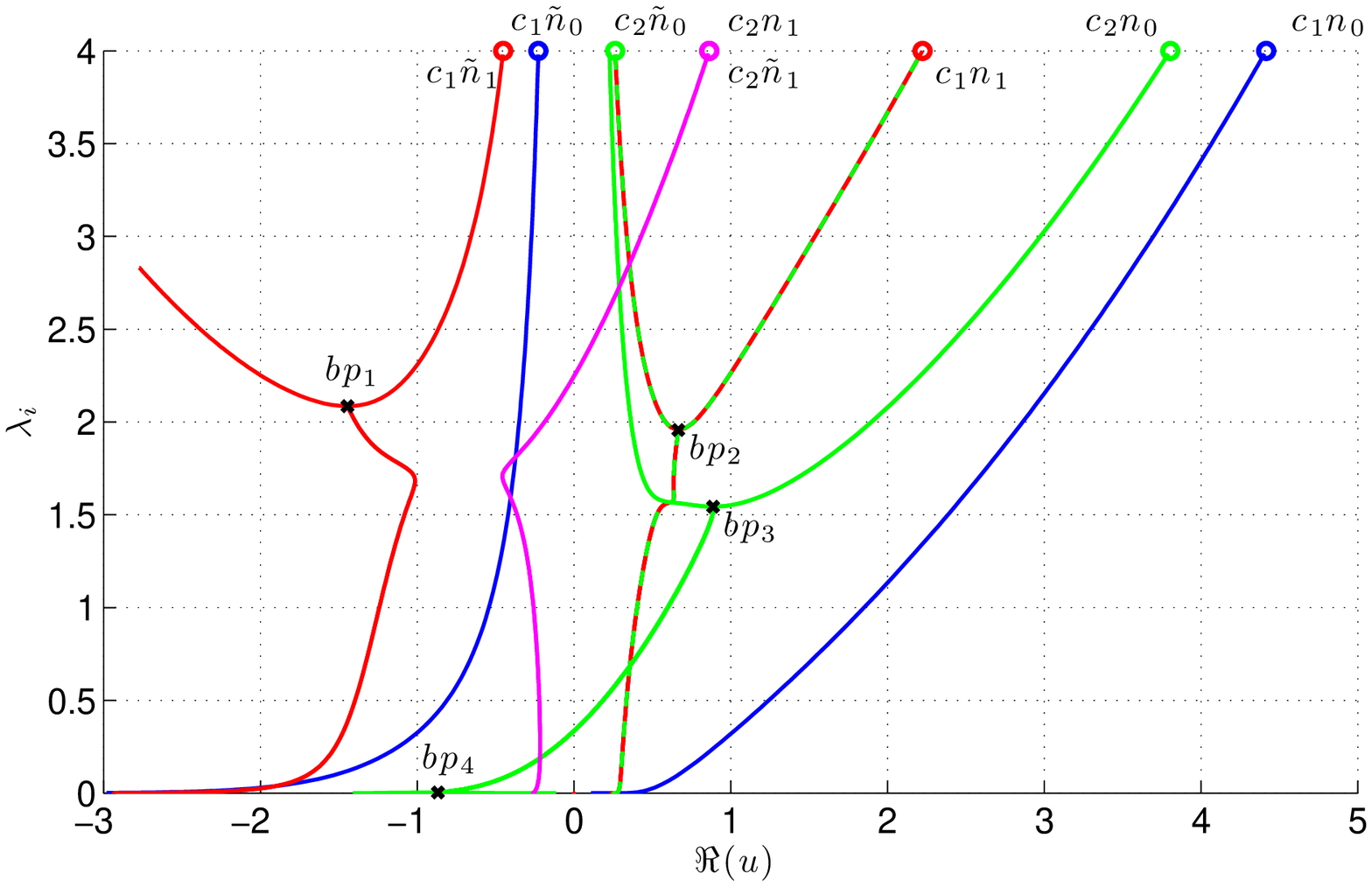}%
		\label{fig:cont_c0.500_re_u}%
	}
	\subfigure[$\Im(u)\times\lambda_{i}$]{%
		\includegraphics[width=4.5cm]{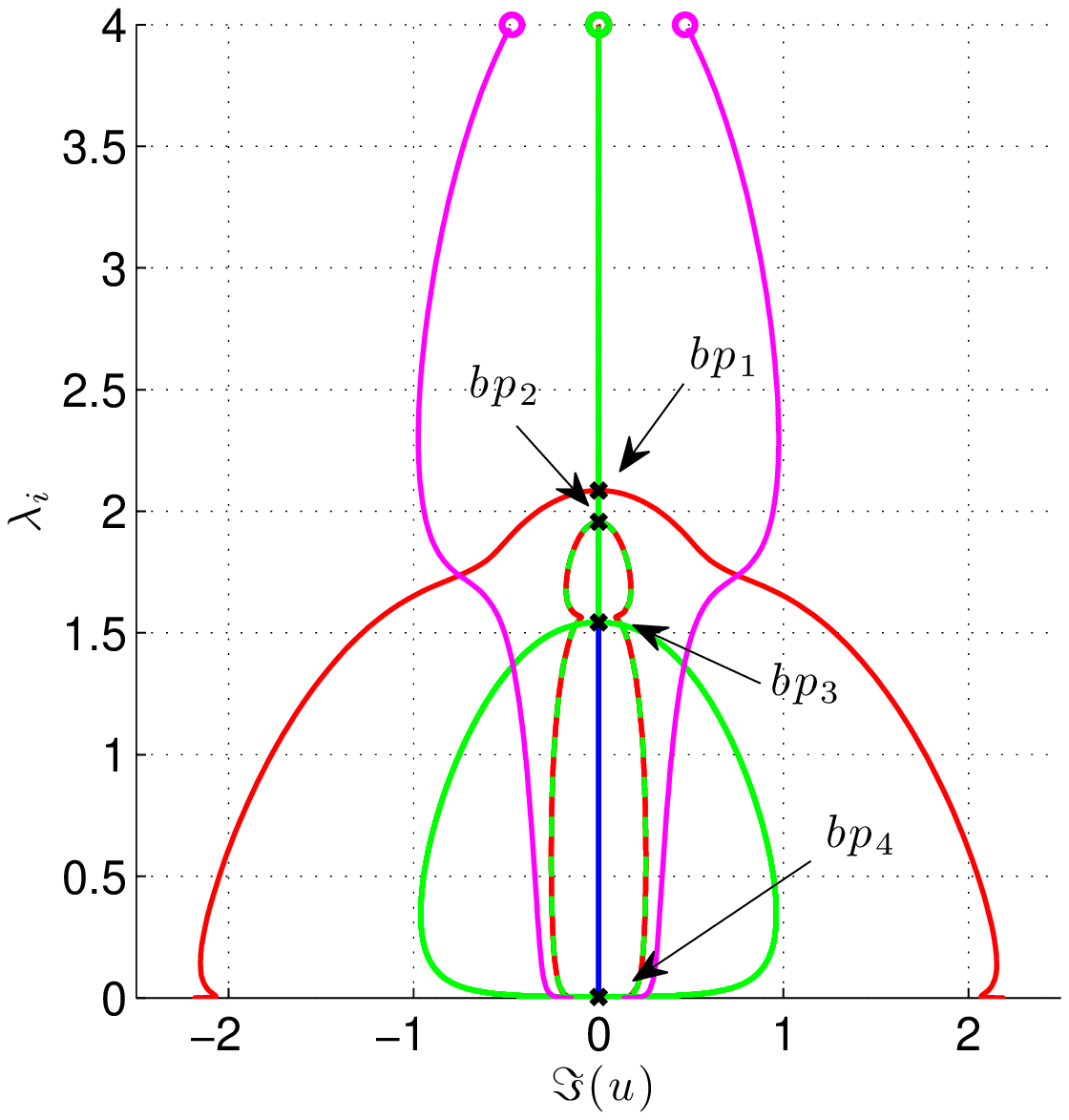}%
		\label{fig:cont_c0.500_im_u}%
	}
	\subfigure[$\Re(u)\times\Im(u)$]{%
		\includegraphics[width=8cm]{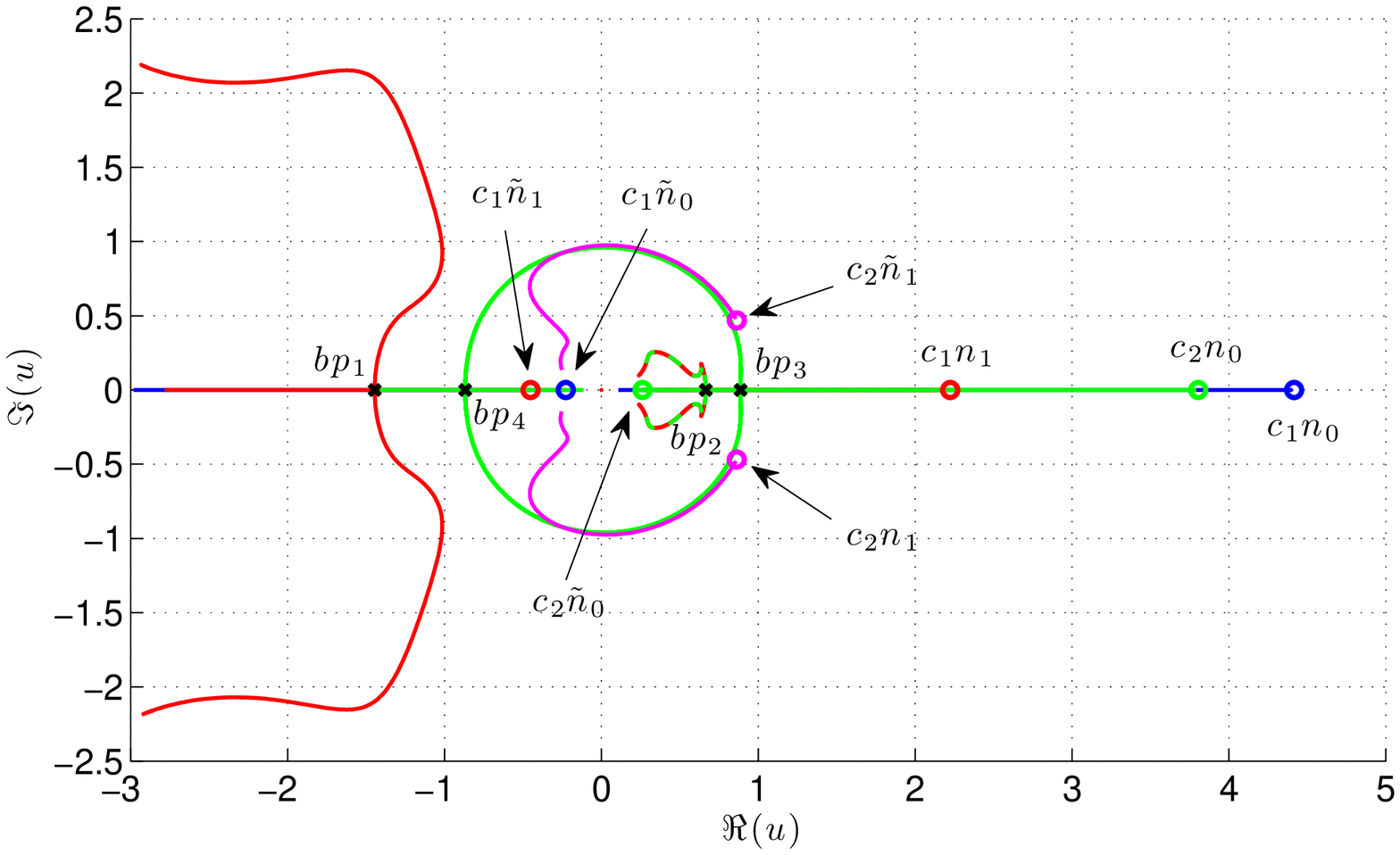}%
		\label{fig:cont_c0.500_reim_u}%
	}
	\hspace{4.5cm}
	\caption{(color online) Different projections of the continuation curves ($\lambda_{c}=0.5$). The four branch points are highlighted and labeled accordingly. Their numerical values are summarized in \tref{tab:021822122010}. The connectivity graph in \fref{fig:co_graph} gives a clearer overview of the associated connections.}
	\label{fig:cont_c0.500_u}
\end{figure}

\begin{figure}[h!]
	\centering
	\includegraphics[width=12cm]{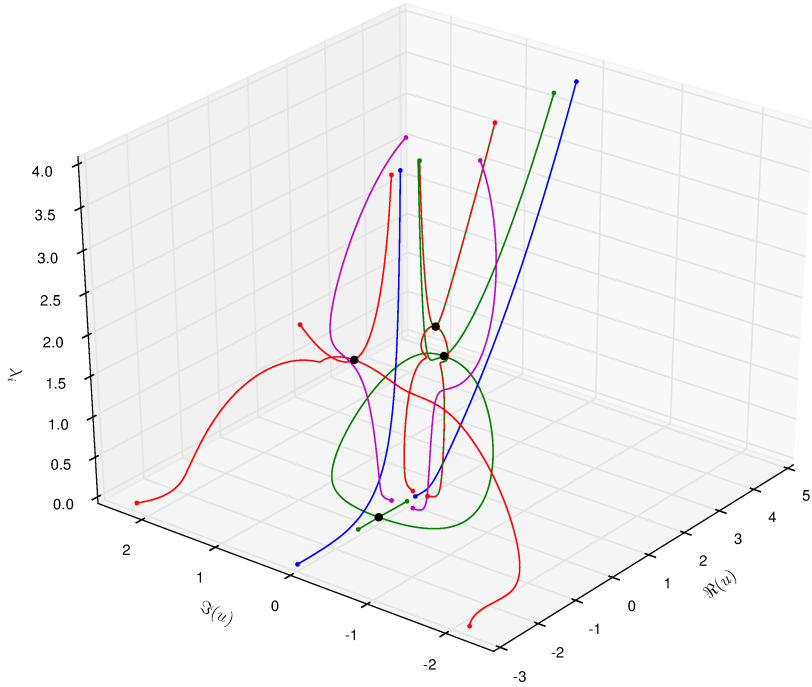}
	\caption{(color online) Continuation curves ($\lambda_{c}=0.5$) shown in the full $\Re(u)\times\Im(u)\times\lambda_{i}$ space. Both start and end points of the curves are highlighted as points in respective colors. The four bifurcation points are indicated with big black dots.}
	\label{fig:cont_c0.500_3d_u}
\end{figure}

\begin{figure}[h!]
	\centering
	\includegraphics[width=5cm]{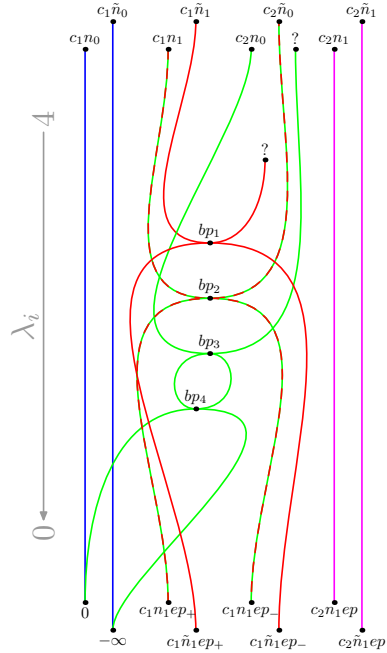}
	\caption{(color online) The connectivity diagram of the different continuation curves. Curves only intersect at the four points $bp_{i}$. Dashed (red/green) line connects two states from different channels through $bp_{2}$. See \fref{fig:closeup_c0.x00} for a more detailed view of the connections. The end points of the curves are labeled accordingly but do not play a major role and their labels are omitted in other figures.}
	\label{fig:co_graph}
\end{figure}

\begin{figure}[h!]
	\centering
	\subfigure[$\Re(E)\times\Im(E)$ projection]{%
		\includegraphics[width=9cm]{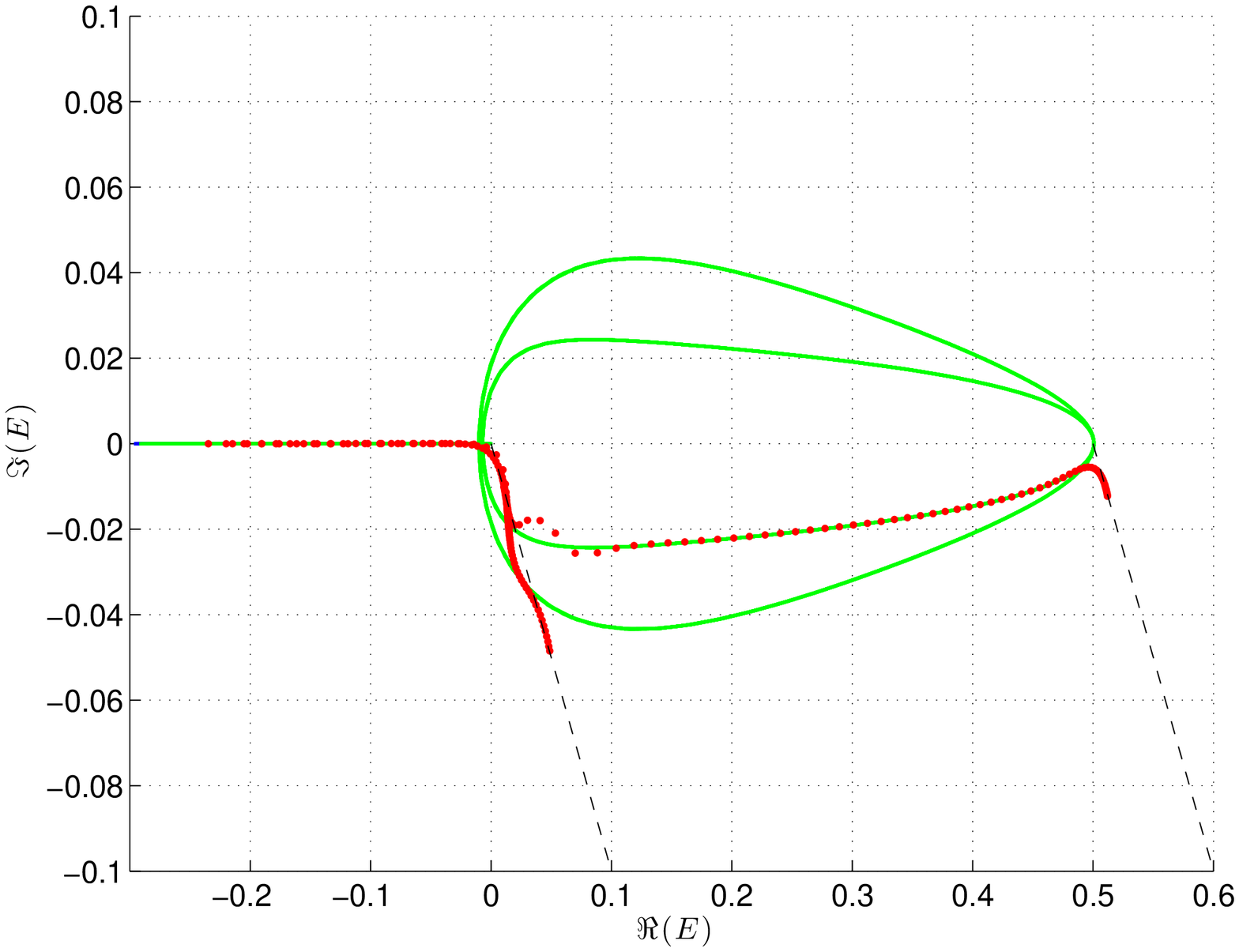}%
		\label{fig:ecs_vs_smatrix_top}%
	}
	\subfigure[$\Re(E)\times\Im(E)\times\lambda_{i}$ space]{%
		\includegraphics[width=9cm]{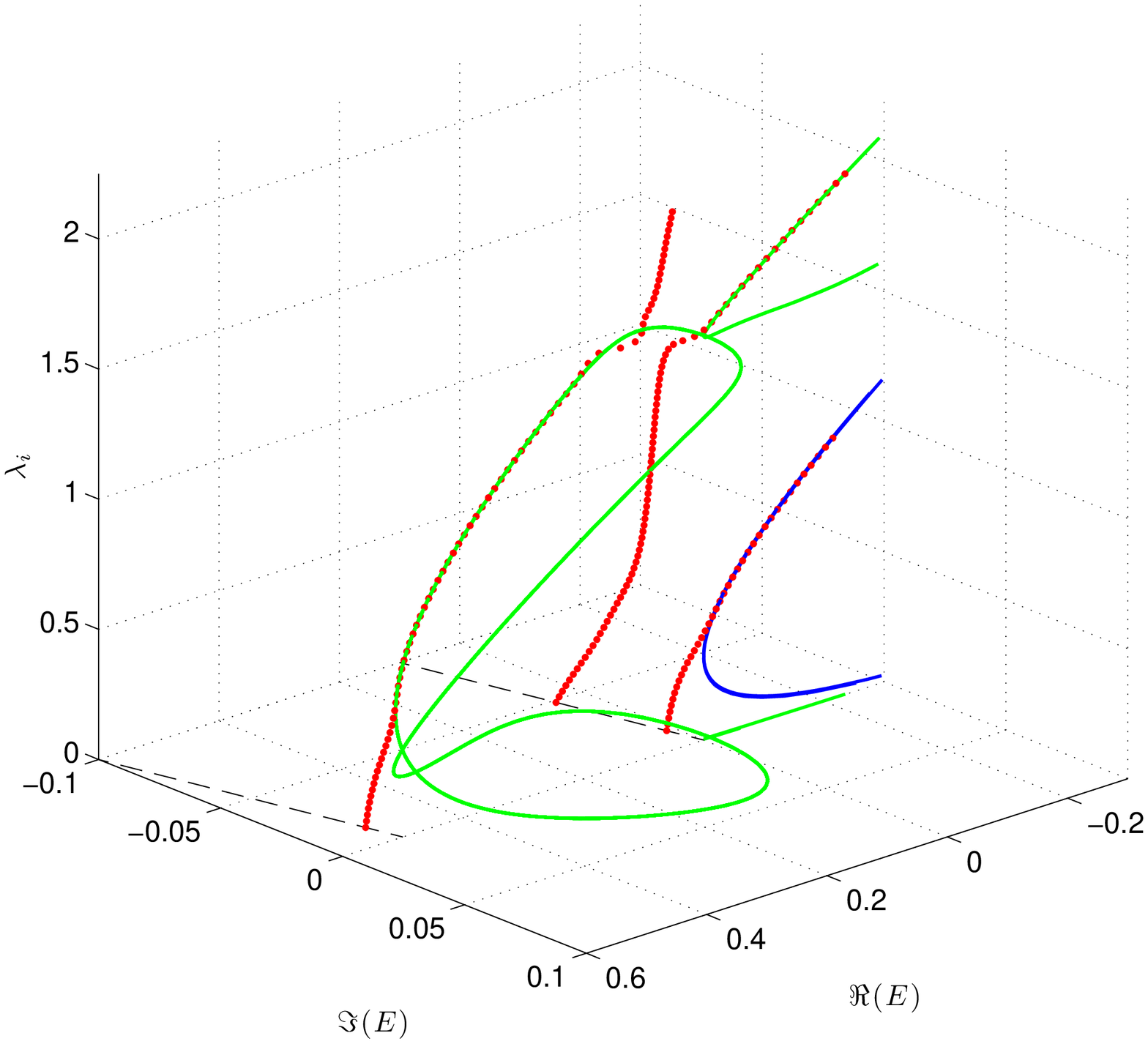}%
		\label{fig:ecs_vs_smatrix_persp}%
	}
	\caption{(color online) A comparison of the results of numerical continuation
          with the results of exterior complex scaling. The vertical
          axis shows the potential strength of the model problem
          while the real and imaginary parts of the energy are shown on the
          other axes. The blue and the green curves are the translation
          of the same curves in the $u$-plane from figures
          \ref{fig:cont_c0.500_u} and \ref{fig:cont_c0.500_3d_u}. The red dots are the relevant ECS eigenvalues calculated for a range of $\lambda_{i}$ values. The results of the two methods
          differ significantly in the regions where a resonance
          becomes a bound state. Also for small potential strengths we
          see significant deviations since ECS cannot resolve the virtual states.}
	\label{fig:ecs_vs_smatrix}
\end{figure}

Although we have performed numerical continuation starting from all states for all subsequent coupling strengths from \tref{tab:095115122010}, we focus on the case $\lambda_{c}=0.5$ as a highlighted example.

The resulting continuation curves are shown in \fref{fig:cont_c0.500_u} as projections on the planes $\Re(u)\times\lambda_{i}$, $\Im(u)\times\lambda_{i}$ and $\Re(u)\times\Im(u)$. Complex geometries and intersections are constructed automatically by the continuation method. An attempt to present those schematically is shown in \fref{fig:co_graph} as a connectivity graph. In the range $\lambda_{i}\in[0,4]$ four branch points can be distinguished with values detailed in \tref{tab:021822122010}. A three-dimensional overview of the continuation curves is shown in \fref{fig:cont_c0.500_3d_u}.

\begin{figure}[h!]
	\centering
	\subfigure[$\lambda_{c}=0$]{%
		\includegraphics[width=8cm]{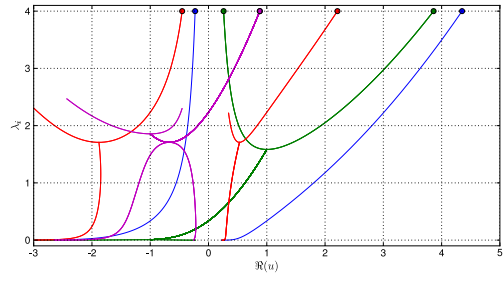}%
		\label{fig:cont_c0.000_side}%
	}
	\subfigure[$\lambda_{c}=0.2$]{%
		\includegraphics[width=8cm]{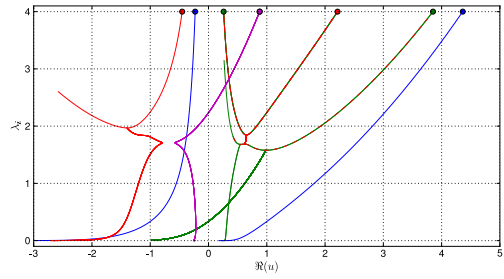}%
		\label{fig:cont_c0.200_side}%
	}
	\subfigure[$\lambda_{c}=0.3$]{%
		\includegraphics[width=8cm]{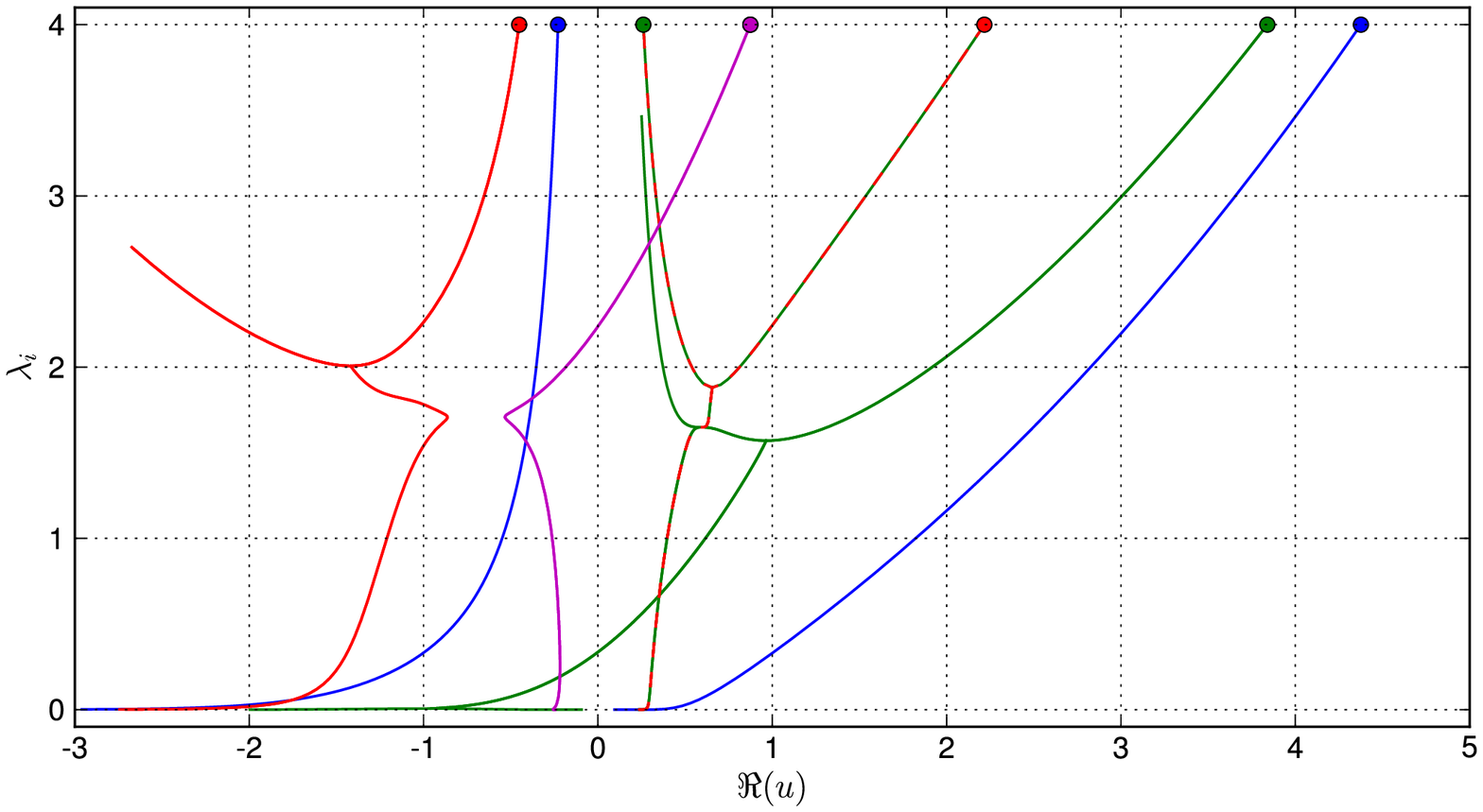}%
		\label{fig:cont_c0.300_side}%
	}
	\subfigure[$\lambda_{c}=0.5$]{%
		\includegraphics[width=8cm]{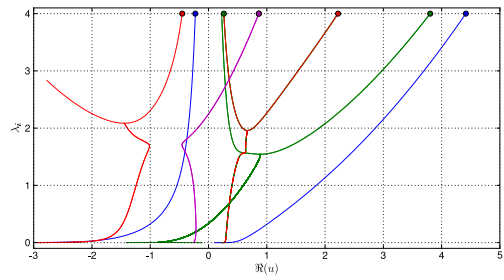}%
		\label{fig:cont_c0.500_side}%
	}
	\caption{(color online) $\Re(u)\times\lambda_{i}$ projections of the continuation curves for different coupling strengths $\lambda_{c}$. Starting values are taken from~\tref{tab:095115122010}.}
	\label{fig:cont_c0.x00_side}
\end{figure}

\begin{figure}[h!]
	\centering
	\subfigure[$\lambda_{c}=0$]{%
		\includegraphics[width=6cm]{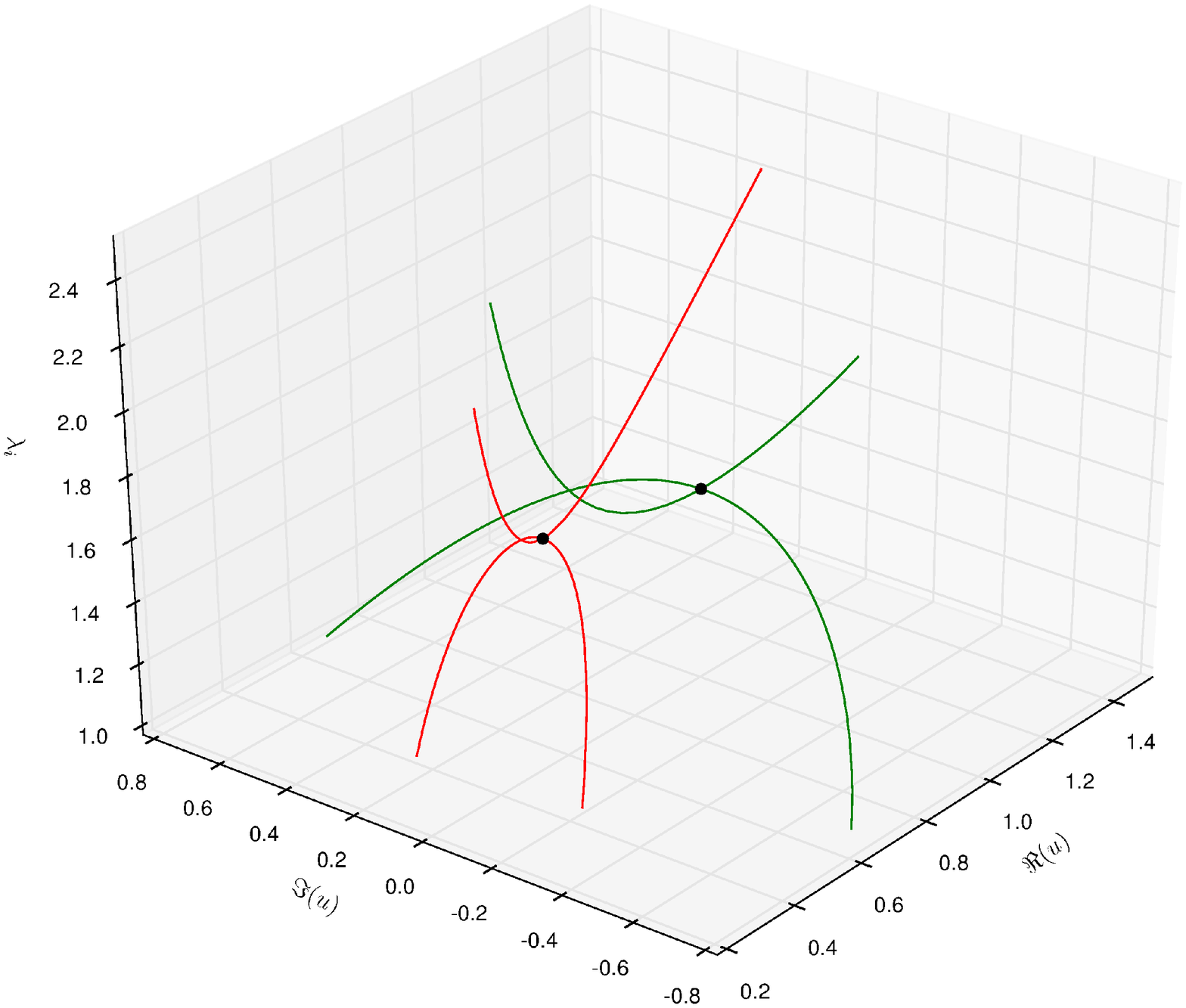}%
		\label{fig:closeup_c0.000}%
	}
	\subfigure[$\lambda_{c}=0.2$]{%
		\includegraphics[width=6cm]{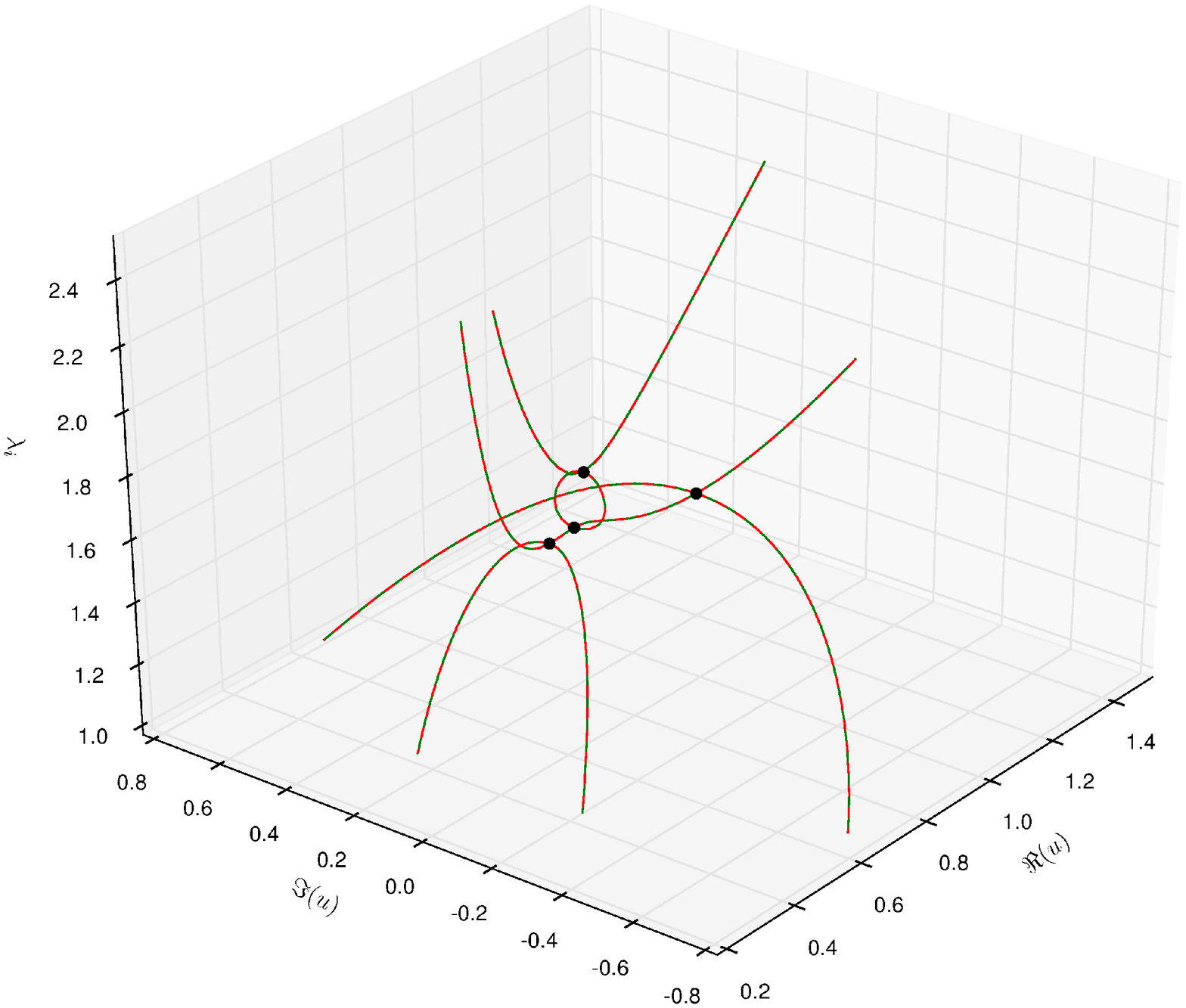}%
		\label{fig:closeup_c0.200}%
	}
	\subfigure[$\lambda_{c}=0.3$]{%
		\includegraphics[width=6cm]{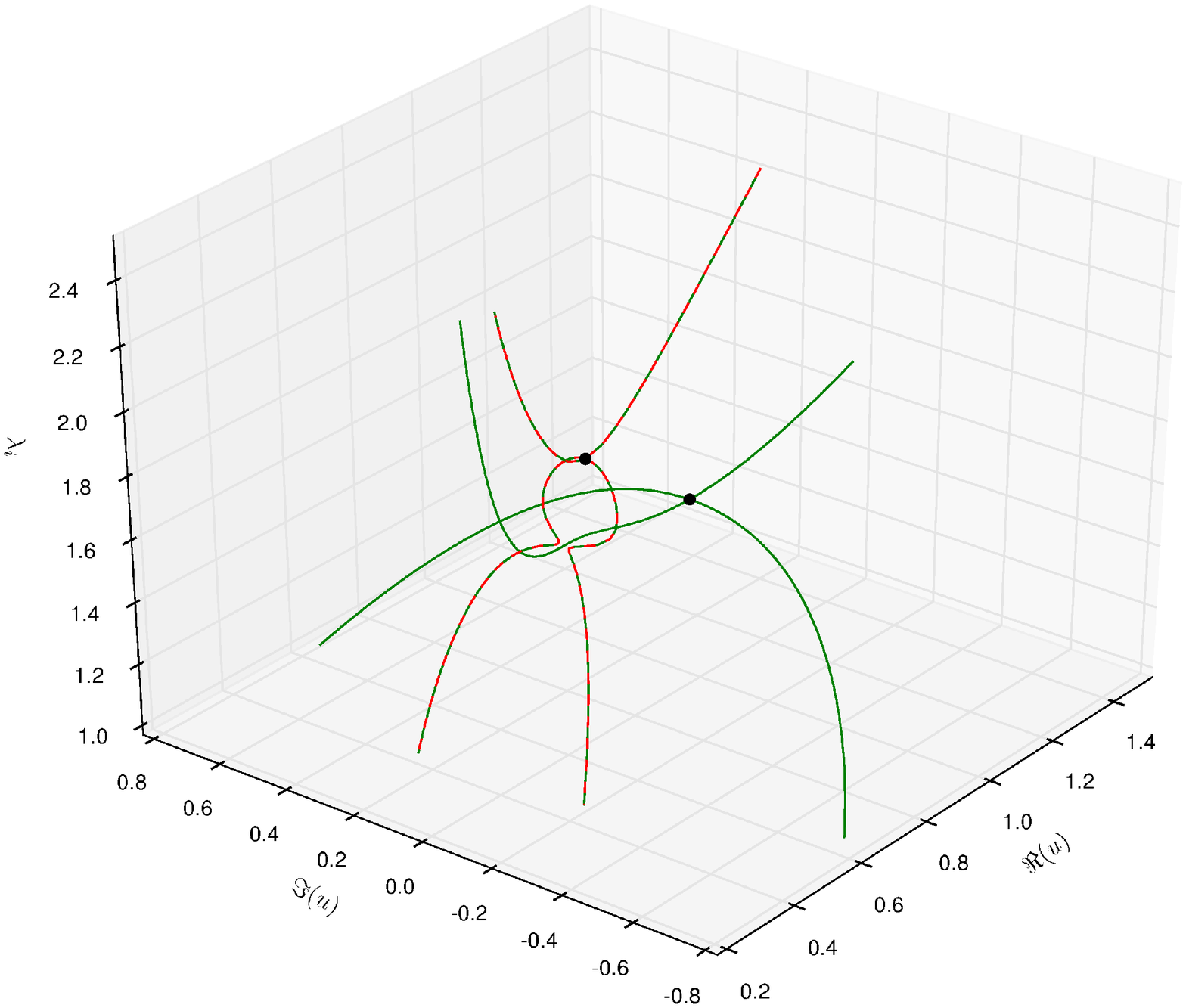}%
		\label{fig:closeup_c0.300}%
	}
	\subfigure[$\lambda_{c}=0.5$]{%
		\includegraphics[width=6cm]{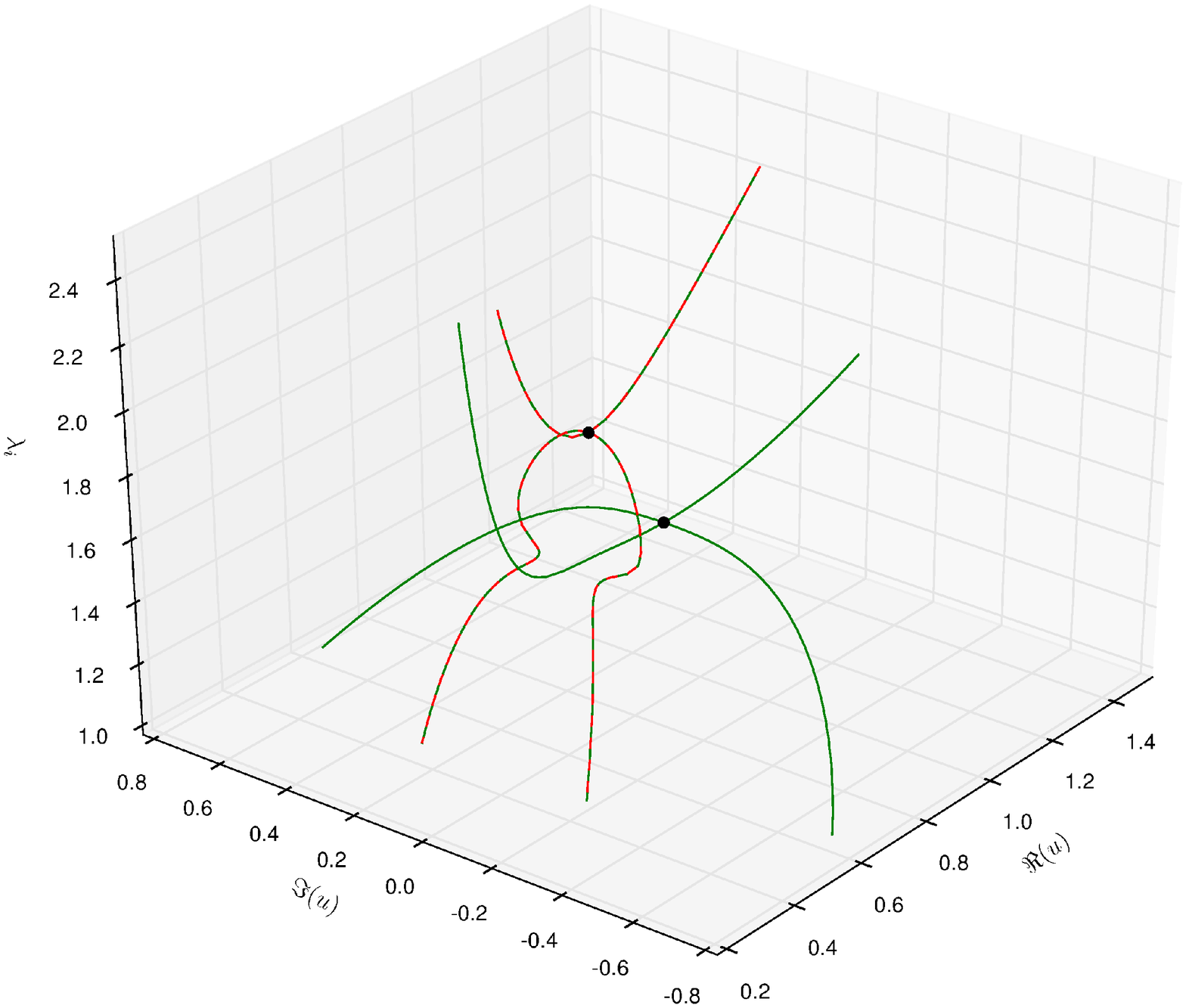}%
		\label{fig:closeup_c0.500}%
	}
	\caption{(color online) Closeup view of the continuation curves of ${\color{green}c_{2}n_{0}}$, ${\color{green}c_{2}\tilde{n}_{0}}$ and ${\color{red}c_{1}n_{1}}$ obtained for different values of the coupling strength $\lambda_{c}$. As the coupling strength increases an interesting effect of reordering of the connections can be observed. Initially the trajectories do not interact. With slight coupling, two additional branch points appear connecting both curves. As the coupling increases, the branch points collide and disappear disconnecting the reordered curves.}
	\label{fig:closeup_c0.x00}
\end{figure}

We have carried out the same procedure for four different values of the coupling strength using starting values summarized in \tref{tab:095115122010}. As the coupling between the two channels increases, resonant trajectories undergo major qualitative changes. See \fref{fig:cont_c0.x00_side} for a short comparison of the continuation paths projected on $\Re(u)\times\lambda_{i}$. The interpretation of those is beyond the scope of this paper although an important effect can be observed. In \fref{fig:closeup_c0.x00} a close-up view of two resonant trajectories is shown. In the uncoupled case (\fref{fig:closeup_c0.000}) two independent trajectories of ${\color{green}c_{2}n_{0}}$ (and ${\color{green}c_{2}\tilde{n}_{0}}$) and ${\color{red}c_{1}n_{1}}$ are shown. Although the green and red curves intersect they do not share a branch point. The addition of a nonzero coupling in \fref{fig:closeup_c0.200} changes this situation drastically: two more bifurcation points appear and the red and green curves are now fully connected (dashed green/red lines). As the coupling strength increases in \fref{fig:closeup_c0.300} the two additional branch points collide and disappear. For an even higher coupling (\fref{fig:closeup_c0.500}) one can clearly see how the curves have rearranged their connections: The ${\color{red}c_{1}n_{1}}$ state is now connected with ${\color{green}c_{2}\tilde{n}_{0}}$ through the dashed green/red line, whereas ${\color{green}c_{2}n_{0}}$ shows a connection with the end point of ${\color{red}c_{1}n_{1}}$ from the uncoupled case.

This short example highlights the ability of continuation methods to deal with subtle and complex connections.

\subsection{Comparison with Exterior Complex Scaling}

To highlight the advantages of the numerical continuation method applied to the function $F(u,\lambda)$, we compare
  its results  with those of a calculation
  with exterior complex scaling (ECS)~\cite{moiseyev1998quantum}. This is done for various
  choices of the parameter $\lambda_{i}$.  First, we translate the curves
  that were obtained by numerical continuation in the $u$-plane back
  to the $E$-plane. These are shown as the green and blue curves in figure
  \ref{fig:ecs_vs_smatrix}. For clarity, the colors are identical to those in the figures depicting the $u$-plane. 
  Note that two different blue curves map to the same region in the $E$-plane. The vertical axis is the strength
  $\lambda_{i}$ of the channel potentials, while the bottom plane shows the complex
  energy of the resonant state. The real part is 
  the proper resonance energy and the imaginary part is the inverse lifetime of
  the resonance. We have found both the exponentially decaying and
  exponentially growing states with a negative, respectively, positive imaginary
  energy.

A first bound state is formed as $\lambda_{i}$ increases and the
  potential becomes stronger. It starts out as a virtual state
  on the negative real energy axis and, with increasing $\lambda_{i}$,  its
  real part increases, goes through zero and becomes negative again. This curve is shown in blue in figure \ref{fig:ecs_vs_smatrix}.

For potential strength $\lambda_{i}$ between $0.33538$ and $1.54368$ we also have
  a resonant state with the real part of the energy between the
  values of the two thresholds. This state is a Feshbach resonance
  related to the second channel that decays through coupling with the first channel, which is open. This state also originates as a virtual
  state on the negative real axis at small potential strengths,
  similar to the state discussed above. However, it is now shifted up
  by $0.5$, the threshold of the second channel. Furthermore, because of
  the coupling to the open channel this virtual state has a
  finite lifetime as soon as it lies above the threshold of the open
  channel. When we increase the potential strength above $1.55204$ this resonance
  becomes a second bound state, but before it becomes bound, there is a bifurcation with a virtual state at parameter strength
  $\lambda_{i}=1.54368$. At this bifurcation point the state has negative real energy.
  As we further increase the strength, one of these states becomes a
  true bound state after passing through zero at potential strength
  $\lambda_{i}=1.55204$. The other point that emerges from the bifurcation
 is a true virtual state that moves down the negative real axis  as a function of $\lambda_{i}$.

When the resonances are calculated by ECS (shown as red dots in figure~\ref{fig:ecs_vs_smatrix}), we do
  not resolve all these details. Especially the connections through
  the virtual states are missing from the picture.  We discretize the
  two channel Hamiltonian on a finite difference grid with grid
  distance $0.03$ and we use an exterior complex scaling transformation that
  starts at $r=12$ and a rotation angle of $\pi/8$. The exterior domain extends to $\Re(r)=15.6$.

As expected, at zero potential strength the eigenvalues of the
  exterior complex scaled Hamiltonian are the discrete eigenvalues of the kinetic
  energy operator. These are related to standing waves on the exterior complex
  scaled domain.  These discrete continuum states will be rotated over
  twice the ECS rotation angle. Therefore, for each threshold we have a
  series of rotated eigenvalues.

As the potential strength increases, these continuum states are
  attracted by the potential and become bound. The first bound state
  is formed from the smoothest continuum state associated with the
  first threshold. It becomes bound at $\lambda_{i}\approx 0.5$.  This is in
  contrast to the numerical continuation results where this bound
  state originates from a virtual state. The difference between the
  numerical continuation and ECS is most prominent at small
  potential strengths.

The second bound state in ECS is formed at potential strength
  $\lambda_{i}\approx 1.57$. Although this state should find its origin at zero potential strength in the smoothest continuum state of the second threshold, an avoided crossing appears instead. The curve originates from the second smoothest state of the first threshold and passes through an avoided crossing with the trajectory of the state that starts from the smoothest state of the second threshold. The former curve partly represents the bound state, whereas the latter curve partly follows the trajectory of the Feshbach resonance between potential strengths $0.46 \lesssim \lambda_{i} \lesssim 1.45$. However, as the curves are disconnected, this calculation misses the bifurcation at $\lambda_{i}=1.54368$ and the subsequent transition through a virtual state. Note that these issues occur near the line of ECS eigenvalues, starting from the first threshold going twice the ECS angle downwards the complex plane. Similary, the deviation of the trajectory around $\lambda_{i}=0.46$, as the Feshbach resonance turns into a virtual state, starts near an analogous line originating from the second threshold. These lines are drawn dashed in figure~\ref{fig:ecs_vs_smatrix}.

\begin{table}
	\footnotesize
	\centering
	\begin{tabular}{|l|rrr|}
		\hline
		Label & $\lambda_{1}=\lambda_{2}$ & $u$ & E \\
		\hline
		$bp_{1}$ & \texttt{2.0852303e+00} & \texttt{-1.4443524e+00} & \texttt{-7.0688100e-02} \\
		$bp_{2}$ & \texttt{1.9571562e+00} & \texttt{6.6636568e-01} & \texttt{-8.7009530e-02} \\
		$bp_{3}$ & \texttt{1.5436785e+00} & \texttt{8.8709701e-01} & \texttt{-7.2105286e-03} \\
		$bp_{4}$ & \texttt{4.0009060e-03} & \texttt{-8.6796523e-01} & \texttt{-1.0092983e-02} \\
		\hline
	\end{tabular}
	\caption{Numerical values of the branch points of the resonant trajectories shown in \fref{fig:cont_c0.500_u}. The coupling strength is $\lambda_{c}=0.5.$}
	\label{tab:021822122010}
\end{table}

\section{Conclusions and Outlook}
\label{sec:conclusion}

An automatic, robust and inherently efficient method for tracking
parameter dependent resonant solutions of the Schr\"odinger equation
is desirable since they play an important role in many quantum
mechanical systems.  Numerical continuation based on pseudo-arclength
continuation has a proven track-record in being reliable and has been
used in the study of various dynamical systems.

We have shown earlier that numerical continuation can be applied to
track bound and resonance states in a single and coupled channel
Schr\"odinger equation with equal thresholds. However, when the
thresholds are different additional branch cuts appear. This leads to
numerical difficulties.

In this article we have shown that these numerical difficulties can
be avoided when a suitable uniformization is applied. The channel momenta (i.e.\ the wave numbers) are
then written as a function of a complex valued parameter.  The
numerical continuation is applied to this parameter combined with
the variable parameters of the problem. Unfortunately, this approach cannot be
extended easily to systems with more than two channels because the
uniformization procedure becomes too complex.

We have applied the method to a two channel problem with Gaussian
potentials and continued in the strength of the potential for various
choices of coupling strength. Several branch points were detected and
the continuation automatically identified the other branches emerging
from these points.  Transitions between bound and resonant states are
easily taken by this method.  In a similar way we could have continued
in the threshold values or any other parameters starting from any solution
point.

The comparison of the results with an exterior complex scaling
  calculation shows significant differences in parameter ranges where the resonance
  transitions to a bound state.  The numerical continuation results predict that
  this transition happens through a virtual state for the model problem. ECS, however, cannot
  resolve these virtual states and the resonance transitions directly to a
  bound state.

In our calculations we have detected several branch points where
  different states meet. All the branch points we have identified, however, are
  bifurcations on the negative real energy axis where two virtual
  states meet.  This occurs when a resonance becomes bound through a
  scenario that was already discussed by Nussenzveig \cite{Nussenzveig1959}.

In the future we will extend the method to higher dimensional problems
with multiple reaction coordinates.

Another possible future direction of research is to use two
parameter continuation and automatically identify the exceptional
points
\cite{heiss2000,mondragon1993degeneracy,vanroose1997double}
where two resonances coalesce at a complex valued energy that does not
necessarely lies on the real axis.
\section*{References}
\bibliography{paper.bbl}

\end{document}